\documentclass[pdftex,twocolumn,epjc3]{svjour3}          % twocolumn

\RequirePackage[T1]{fontenc}

\usepackage{amssymb}
\usepackage{amsmath}
\usepackage{bm}
\usepackage[dvipsnames]{xcolor}
\DeclareOldFontCommand{\bf}{\normalfont\bfseries}{\mathbf}
\usepackage{mathptmx}

\usepackage[normalem]{ulem}

\usepackage{hyperref}
\usepackage{graphicx}
\usepackage{etoolbox}
\usepackage{tikz}
\newrobustcmd*{\mysquare}[1]{\tikz{\filldraw[draw=#1,fill=#1] (0,0)
rectangle (0.2cm,0.2cm);}}
\journalname{Eur. Phys. J. E}

\begin{document}

\title{Hydrodynamic effects on the liquid-hexatic transition of active colloids }

\author{ G. Negro\thanksref{addr1} \and C. B. Caporusso\thanksref{addr1} \and P. Digregorio\thanksref{addr2} \and G. Gonnella\thanksref{addr1} \and A. Lamura\thanksref{addr3} \and  A. Suma\thanksref{addr1} }

\institute{
Dipartimento di Fisica, Università degli Studi di Bari and INFN, Sezione di Bari, via Amendola 173, Bari, I-70126, Italy\label{addr1} \and
Centre  Européen  de  Calcul  Atomique  et  Moléculaire (CECAM), Ecole  Polytechnique  F\'ed\'erale  de  Lausanne (EPFL),  Batochimie,  Avenue  Forel  2,  1015  Lausanne,  Switzerland \label{addr2} \and
Istituto Applicazioni Calcolo, CNR, Via Amendola 122/D, I-70126 Bari, Italy\label{addr3}
}

\date{Received: date / Accepted: date}

\abstractdc{
We  study numerically the role of hydrodynamics in the liquid-hexatic transition of active colloids at intermediate activity, where motility induced phase separation (MIPS) does not occur.
We show that in the case of active Brownian particles (ABP), the critical density of the transition decreases upon increasing the particle’s mass, enhancing ordering, while self-propulsion has the opposite effect in the activity regime considered. 
 Active hydrodynamic particles (AHP), instead,  undergo the   liquid-hexatic transition  at  higher values of packing fraction $\phi$ than the corresponding ABP,  suggesting that hydrodynamics have the net effect of disordering the system. At increasing densities, close to the hexatic-liquid transition, we found in the case of AHP the appearance of self-sustained organized motion with clusters of particles moving coherently.}
%\textcolor{black}{to check at the end: suffix s and c are coherent and in the figures. Figures have the label a-b-c-d if indicated in the text. Mettere ref in bibtex, controllare omogeneita sec, Sec, ref o Ref come sono scritte. Controllare captions. Decidere come chiamare trans flow e active time}
\maketitle

\section{Introduction}

%\lino{???Ref a kraft???}

%\textcolor{black}{Aggiungere ref buttinoni lowen, ref a kraft, controllare refs che ci hanno citato (perlomeno prl e aggiungere quelle attinenti. cercare di sep ref se si citano argomenti sperimentali diversi)}
Self-propelled particles (SPP) are the fundamental units of a broad class of theoretical models for active matter. In the context of  SPP models, injected energy from the environment fuels a persistent motion of the single constituents, driving the system out of thermal equilibrium. Simplified models of SPP \cite{filyABPs,romanczuk2012,speckCRYSTALSPP,bechinger-revmodphys}  are of crucial importance, because they offer a minimal setup to explore some of the large variety of collective behaviours observed in nature for systems of motile living bodies at different length scales, from flocking of birds and fish~\cite{bialekFLOCKS}, to swarming in bacterial colonies~\cite{beerBACTERIA} and dynamics in cells' cytoskeleton~\cite{dogicCELLS}.

Active Brownian Particles (ABP) models are  very popular among SPP models~\cite{nostroDISCHI,claudioDISCHI,filyABPs}. Active colloids are  usually spherical particles undergoing directed motion due to an active force, while both translational and rotational degrees of freedom are in contact with a stochastic thermal bath. Although the model is very simple, ABP show paradigmatic collective phenomena like motility-induced phase separation (MIPS)~\cite{catesMIPS,filyABPs,gonnellaMIPS,rednerMIPS,buttSPP} and are therefore very interesting in order to characterize the fundamental principles governing active matter systems. Moreover, ABP are of primary use for comparisons with experimental systems of synthetic micro-swimmers~\cite{buttSPP,demianSPP}, opening the perspectives for a systematic control of active systems and collective motion, with the purpose to exploit some of their unique features for technological uses, for instance in robotics~\cite{robotswarm,smarticles,robotlocomotion}, realisation of biological machines~\cite{kapral_nanomotors}, or understanding of flocking intelligence~\cite{PNAS_schooling,biderale_swimming-strategies}.

%Numerous specific models have been devised, starting from the key ideas behind ABPs, that allow to reproduce many of these phenomena by means of numerical simulations and continuous theories. Moreover, 
%\textcolor{black}{Separa ref per argomento}
Of particular interest is the characterization of ABP in the dense regime,  see e.g. spontaneous flow~\cite{yeomansDENSE} or glassy behaviour~\cite{berthierDENSE,mandalDENSE} in biological tissues, biofilms, cell mono-layers~\cite{henkesDENSE,biACTGLASS}, and can be considered a target
for the development of new materials \cite{smart_materials_1}. In two dimensions (2D), ABP present ordering phase transitions when the density of the system is increased~\cite{krauthACTIVE,nostroDISCHI,claudioDISCHI,nostroDIFETTI}, which are connected to those encountered for passive hard colloids \cite{krauthPASSIVE1,krauthPASSIVE2,thorneyworkACTIVEMELTING}. At intermediate values of the self-propelling force, a liquid-hexatic critical transition is followed by a hexatic-solid transition, where the solid phase has quasi-long-range (QLR) positional and long-range (LR) orientational order, the hexatic phase has short-range (SR) positional and QLR orientational order, while the liquid phase is homogeneous and has SR positional and orientational order. This scenario is very similar to the theoretical Kosterlitz, Thouless, Halperin, Nelson, and Young (KTHNY) two-step scenario\cite{KT_1,KT_2,KT_3}. If activity is high enough, instead, MIPS takes place, as a phase separation between a dense phase and a gaseous one~\cite{nostroDISCHI}.

The aforementioned features of the ABP phase diagram have been well established in the context of over-damped motion and without an explicit underlying thermo-hydrodynamic bath. At the same time, there are other interesting questions that remain to be considered.  %several questions still remain unresolved. % about the effect  but little is know about their effects in the dense regime at intermediate activities, and how they modify the KTHNY scenario. Here we raise here two points that need further investigations. 
The first question concerns the role of \textcolor{black}{particles mass}, and in particular the interplay between inertial and active diffusion timescales, which can be varied independently~\cite{pica2020,capriniINERTIA}. 
It has been pointed out in~\cite{omar2021tuning} that in three-dimensional active systems, inertia should attenuate the destabilizing effect of activity on the ordered phase. %\lino{What about~\cite{liaoINERTIA}? Shell we mention it later?}.
The presence of large inertia has also been shown to strongly affect the kinetic energy of the particles into the highly dense phase of MIPS~\cite{isabellaTEFF}, and to highly inhibit phase segregation~\cite{lowenINERTIA}. However, the role of inertia in the context of dense ABP, and in particular how the particle's mass affects the hexatic phase, has not yet been characterized.  
The second question concerns the role of hydrodynamic interactions in the dense phase. Regarding the influence of hydrodynamics in MIPS, it is found that in 2D MIPS is suppressed~\cite{theersSRD,navarro2014,gompperROADMAP}, as hydrodynamics favour reorientation of particles’ self propulsion direction, while in quasi-2D systems MIPS has been observed for low-density fluids\cite{stark} and not when the fluid was made incompressible~\cite{thee2018,gompperROADMAP}. For elongated colloids, steric alignment and hydrodynamics show highly non-trivial interplay, such that MIPS is enhanced for pullers and suppressed for pushers~\cite{theersSRD}. 
%As a general rule, alignment interactions systematically compete with self-propulsion, leading to a wide spectrum of collective behaviours~\cite{chateDADAM,peruaniRODS}, both when they arise from excluded volume and anisotropic shape, or when they are explicitly coded into the dynamics, like in the Vicsek model. 

As a first step in the direction of answering these two questions, we characterize how the critical density for the liquid-hexatic transition of active particles is modified, in an intermediate activity regime where MIPS does not occur for ABP, by i) the inertial effects due to mass changes, and ii) the presence of non-isotropic interactions between colloids introduced by hydrodynamics. Hydrodynamics has been implemented by using the multi-particle  collision  method~\cite{male1999,gommper_review_2009}, which seamlessly integrate with the dynamics of active Brownian particles~\cite{plipton_colloids_lammps}.  In particular, we implement thermal slip boundary conditions, decoupling colloids rotational diffusion from the solvent and test the consistency of this implementation with known benchmark tests.
We focus here only on 2D systems where the rotational diffusion follows the same equations as for ABP. This allows us to have an active hydrodynamic particle (AHP) model with the same friction, temperature and rotational diffusion as the ABP model, providing a way to quantitatively compare them.

We find that changing the colloids mass and introducing hydrodynamic interactions affect the critical density at which the liquid-hexatic transition occurs. In particular, mass changes lower this density with respect to over-damped ABP, while hydrodynamics increases the critical density.
 We also find that the system with hydrodynamics undergoes a transition from a disorganized to a self-sustained flow regime upon increasing the density, with particles moving on the same direction at high densities.

The work is organized in the following way. In Sec. II we discuss the numerical methods and parameter choice for the ABP model and for  the AHP model, with Sec. II C providing several tests for implementation of the latter model. In Sec. III A we discuss how the liquid-hexatic scenario changes by \textcolor{black}{varying the active colloids mass}, while in Sec. III B we discuss the effects due to hydrodynamics interactions. 
Finally we draw some conclusions discussing the main findings.

\section{Numerical methods}
\label{sec:model}
In this Section we describe the numerical models. We will start with the ABP model, which follows a Langevin equation and does not include hydrodynamic interactions. We will then describe the AHP model, where hydrodynamic is accounted explicitly, and provide some numerical tests of the implementation.

\subsection{Active Brownian particles (ABP)} 
\label{sec:colloids}
We consider a two-dimensional system with $N_c$ disks of mass $m_c$ and diameter $\sigma_c$ in a square box size of side $L$. Each disk $i$ has also an associated axis $\mathbf{n}_i=( \cos{\theta_i(t)},\sin{\theta_i(t)})$, where $\theta_i$ is the angle between the axis and the x-axis and which evolves over time. $\mathbf{n}_i$ represents the direction in which the self-propulsion occurs. 

The particles interact with each other 
via a short-ranged repulsive potential:
\begin{equation}
U(r)= 4 \epsilon \Big [ \Big ( \frac{\sigma}{r} \Big )^{64}
                      - \Big ( \frac{\sigma}{r} \Big )^{32}
                      + \frac{1}{4}
                 \Big ]\Theta(\sigma_c-r)
\label{eq:pot}
\end{equation}
where $r$ is the inter-particle distance between the center of masses of each colloid, $\Theta(r)$ is
the Heaviside function ($\Theta(r)=0$ for $r<0$ and 
$\Theta(r)=1$ for $r \geq 0$), and $\sigma=2^{-1/32}\sigma_c$. 

The evolution of the centre of mass of disks is described by a Langevin equation, with activity modelled as a  force $F_{\rm act}$ of constant magnitude  acting along the particle axis $\mathbf{n}_i$, while the propulsion axis changes its direction in time through a diffusion equation:
\begin{gather}
    m_c\Ddot{\bm{r}_i} = - \gamma \dot{\bm{r}_i} + F_{\text{act}}\mathbf{n}_i - \bm{\nabla_i}\sum_{j\neq i}U(r_{ij}) + \bm{\xi}_i, \label{eq:langevin} \\
    \dot {\theta}_i  = \eta_i \ ,
    \label{eq:diffusion}
\end{gather}
where $i=1,..,N_c$, $r_{ij}=|\mathbf{r}_i-\mathbf{r}_j|$ and $\gamma$ is the damping coefficient. The terms $\bm{\xi_i}$ and $\eta_i$ are Gaussian white noises that mimic the interaction with a thermal bath, with average zero and  variance  fixed by the fluctuation-dissipation theorem:
\begin{gather}
    \langle \xi_{i\alpha}(t) \rangle = 0,\; \langle \eta_{i}(t) \rangle = 0, \\
    \langle \xi_{i\alpha}(t_1) \xi_{j\beta}(t_2) \rangle = 2k_B T \gamma \delta_{ij}\delta_{\alpha\beta}\delta(t_1-t_2),\\
    \langle \eta_{i}(t_1) \eta_{j}(t_2) \rangle = 2D_\theta \delta_{ij}\delta(t_1-t_2),
\end{gather}
where $\alpha,\beta=1,2$ are the indices of the spatial coordinates, $T$ the temperature of the system,  $k_B$ the Boltzmann constant and $D_\theta$ the rotational diffusion coefficient.  
We express all the quantities in units of mass, length and energy ($\tilde{m}$, $ \tilde{\sigma}$ and $\varepsilon$, respectively), with the time unit expressed as  
 $\tau = (\tilde{m}{\tilde{\sigma}}^2/\epsilon)^{1/2}$.  Note that we fix $\sigma_c=1\tilde\sigma$, while $m_c$ is varied with respect to the mass unit $\tilde{m}$. From now on we will drop the units for simplicity.
 
The density of the system is expressed in terms of the packing fraction $\phi =\pi{\sigma^2_c}N_c/(4L^2)$, ratio between the surface occupied by the colloids and the total system surface $L^2$. An important adimensional number, which measures the ratio between the active work required to move a particle by $\sigma_c$ and the typical thermal energy $k_BT$, is the P\'eclet number Pe = $F_{\rm act} {\sigma_c}/(k_BT)$. 
Another useful adimensional number is the active  Reynolds number, which measures the ratio between inertial and viscous forces \textcolor{black}{acting on the colloids}, $Re_{act}=\frac{m_c F_a}{\sigma_c\gamma^2}$ \cite{PhysRevE.90.052130}.

The typical  time scales for a single ABP are the inertial time $t_I=m_c/\gamma$ and the persistence time $t_p=1/D_{\theta}$, with the latter signaling the crossover to the final diffusive regime and  that depends only on the rate of rotational diffusion and not on the activity parameter. We can define a useful adimensional number as the ratio between $t_I$ and $t_p$, to which we refer to hereafter as the persistence number $pn=t_I/t_p$.

We fix in our numerical simulations $\gamma=10$, as previously done for ABP \cite{nostroDISCHI} where the choice $m_c=1$ was adopted, which corresponds to limit inertial effects at small times $t_I=0.1$.   In the following we keep fixed $\gamma$ and vary the disk mass $m_c$ to consider different inertial contributions, $k_BT=0.05$ and $D_\theta = 3 k_BT/(\sigma^2_c \gamma)=0.015$. We  fix  $N_c=16384$ and vary $L$ in order to obtain the correct packing fraction $\phi$. We use LAMMPS\cite{plim1995} to integrate numerically the equations of motion, using a timestep  $\Delta t_c=0.001$ and periodic boundary conditions. We fix the P\'eclet number to $\textit{Pe}=$ $5$, $10$ and $20$, and vary the packing fraction $\phi$ between $0.60$ and $0.88$.
Within the range of chosen parameters, $Re_{act}$ is always smaller than one.
For each set of parameters a single realization was considered which was run  between  $10^4$ and $10^5$ simulation time units after steady state was reached. In this time frame averaged quantities were computed. 

\subsection{Active hydrodynamics particles (AHP)}
\label{sec:AHP}
The ABP model described beforehand does not account explicitly for the solvent. In order to add this effect, we choose as model a mesoscopic method known as multi-particle collision (MPC) dynamics, first introduced in \cite{male1999}. After briefly describing the MPC model, we will introduce two possible ways to couple solvent and disks, their dynamics and the specific parameters used for simulations. Tests of this implementation are presented in section \ref{sec:validation}.
 
\subsubsection{Solvent dynamics} 

The solvent consists of $N_s$ identical
point-like particles of mass $m_s$  embedded in a two-dimensional square box of size $L$.
Each particle $i$ is characterized by a position ${\bf r}_i$ and velocity ${\bf v}_i$, both of which are continuous variables. In this algorithm, the time is discretized in units $\Delta t_s$, and the evolution of the system is composed by two steps, propagation and collision, which are applied consecutively for each $\Delta t_s$.

In the
propagation step, particles are freely streamed according to their velocities as
\begin{equation}
{\bf r}_i(t+\Delta t_s)={\bf r}_i(t)+{\bf v}_i(t) \Delta t_s .
\label{eq:prop}
\end{equation}
In order to perform the collision step, the system is partitioned 
into cells of a square lattice with mesh size $\sigma_s$. 
Each cell is the scattering area where a MPC occurs, which updates particles velocities according to the
rule \cite{male1999,male2000}
\begin{equation}
{\bf v}_i(t+\Delta t_s)={\bf u}(t)+\Omega [{\bf v}_i(t)-{\bf u}(t)] ,
\label{eq:coll}
\end{equation}
where ${\bf u}=(\sum_{i=1}^{m} {\bf v}_i)/m$ is the mean velocity of the $m$ 
colliding particles in the cell, also assumed to be  
the macroscopic velocity of the fluid. $\Omega$ is a  rotation 
matrix with  angle $\pm \alpha$ ($0<\alpha<\pi$). The angle $\alpha$ is fixed at the beginning of the simulation while its sign is assigned
with equal probability to every cell at each time step. 
In each cell  all the $m$ relative velocities are rotated with the
same  angle. 
Linear momentum and kinetic energy are conserved under this dynamics. 

The transport coefficients of this model can be analytically derived. In particular, for our purposes the kinematic viscosity $\nu_s$ and the self-diffusion coefficient $D_s$ will be useful.
In 2D the viscosity is equal to\cite{kiku2003,ihle2004}:
\begin{equation}
\begin{split}
\nu_s &=\frac{\sigma_s^2}{2 \Delta t_s}
\Big [ 
\left ( \frac{\lambda}{\sigma_s} \right )^2 
\Big ( \frac{n_s}{(n_s-1+\exp{(-n_s)})\sin^2(\alpha)}-1 \Big )\\
&+\frac{(n_s-1+\exp{(-n_s)})(1-\cos(\alpha))}{6 n_s} 
\Big ],
\end{split}
\label{eq:visc}
\end{equation}
while the coefficient $D_s$ is~\cite{tuze2006}:
\begin{equation}
D_s=\frac{\lambda^2}{2 \Delta t_s}
\Big ( \frac{2n_s}{(n_s-1+\exp{(-n_s)})(1-\cos(\alpha))} \Big ),
\label{eq:diff}
\end{equation}
where $n_s=N_s \sigma_s^2 / L^2$ is the average number of particles per cell
and $\lambda=\Delta t_s \sqrt{k_B T/m_s}$  is
the mean-free path.

\subsubsection{Solvent-colloids coupling}
\label{sec:solvent_colloids}
The next step would be to integrate the solvent particles with the colloids, which means that we need to decide how to couple colloids and solvent dynamics.
Different strategies are possible and a review for MPC with passive colloids can be found in \cite{gommper_review_2009}; here we adopted the one implemented in the LAMMPS software\cite{plipton_colloids_lammps}. 
%\textcolor{black}{\sout{It has been then customized to our system of active particles as described in the following.}A: a che serve sta frase? la toglierei. Il metodo rimane indipendente dall'aver messo la forza attiva}

In this implementation colloids are evolved for $n$ time\-steps $\Delta t_c$, following the equation of motion (\ref{eq:langevin}) without the force terms $\bm{\xi}_i$ and $\gamma \dot{ \bm {r}_i}$, which accounted implicitly for the thermal bath in the ABP model, and are substituted here by the MPC bath.  Afterwards solvent particles are propagated for a timestep equal to $\Delta t_s=n\Delta t_c$.  Note that both $\Delta t_c$ and $\Delta t_s$ are expressed  in the same time unit as in the ABP model. Before computing the collision (\ref{eq:coll}), the algorithm checks if solvent particles are \textcolor{black}{overlapping} with disks having diameter $\sigma_c$ and mass $m_c$, that is if the position of point-like solvent particles is inside the disks area.  In this case, an exchange of momentum occurs, followed by a change in the position of solvent particles to place them
out of the colloids, and, finally, the collision step for solvent particles is applied.
 
The exchange of momentum is decided by the proper colloid–solvent boundary condition (BC) adopted, which can be either no-slip or slip. No-slip BC means that both linear and angular
momentum are exchanged between colloid and solvent particles\cite{bocq1994}, while for slip BC only linear momentum is transferred as in the case with radial interactions \cite{male2000}. Several implementations of the BC are available, such as the so called thermal BC \cite{inou2002} and the bounce-back collision rule\cite{lamu2001}. The latter, used in the case of no-slip BC, requires the use of phantom particles inside the colloid while the former does not. Here we choose the thermal BC method, described below for the slip and no-slip cases, as it is in general
useful under forced flow conditions, like the case of active particles, and is particularly suited when the solvent mean free path is  
much smaller than the disk radius \cite{padd2005,padd2006}. %, although other methods, such as the , which inverts the components of the solvent velocity relative to the surface of collision, could be implemented.
%so that
%the tangential fluid velocity at the 
%surface is equal to the local velocity of the colloid surface
%, as usual for spherical colloidal particles with radial interactions, 
%We use the so called
 
%The LAMMPS implementation~\cite{hech2005} of momentum exchange 

%The exchange of momentum between solvent particles 
%and disks occurs in the following way.

In the no-slip thermal BC, when a solvent particle of velocity ${\bf v}$ overlaps with a disk, it is moved back to the disk surface along the shortest vector ${\bf r}_d$
and then streamed for a distance ${\bf v}' \Delta t_s \varepsilon$, 
where ${\bf v}'$ is the updated velocity 
and $\varepsilon$ is a uniformly distributed random number in the
interval $[0,1]$ \cite{hech2005}.
The new velocity ${\bf v}'$ is divided in the normal $v_N$ and tangential
$v_T$ velocity components with respect to the particle-colloid distance, and chosen according to the stochastic distributions \begin{eqnarray}
p_N(v_N)&=&(m_s v_N/k_B T) \exp{(-m_s v_N^2/2k_B T)} , v_N>0 \label{pn} \\
p_T(v_T)&=&\sqrt{m_s/2\pi k_B T} \exp{(-m_s v_T^2/2k_B T)} ,
\label{dist}
\end{eqnarray}
 centred around the local velocity  ${\bf v}_d$ of the colloid surface, where ${\bf v}_d= {\bf V} + {\bf \omega} \times ({\bf r}_d-{\bf R})$,
with ${\bf R}$ being the position of the colloid centre, ${\bf V}$ and ${\bf \omega}$ 
the linear and angular velocities of the colloid. %\textcolor{black}{non è chiarissimo che vuole dire centered sulle distribuzioni. sarebbe implicitamente $v_N=v_N^{'}-v_{dN}$ ad esempio nella distribuzione?}
Regarding the change in momentum for the colloid after the collision, all the linear and angular momenta variations of the overlapping solvent particles are summed up as
$\Delta {\bf P}=\sum_s m_s ({\bf v}-{\bf v}')$ and 
$\Delta {\bf L}=\sum_s m_s ({\bf r}_d-{\bf R}) \times ({\bf v}-{\bf v}')$, and
 the linear and angular velocities of the colloid are updated as:
${\bf V}'={\bf V}+\Delta {\bf P}/m_c$ and 
${\bf \omega}'={\bf \omega}+\Delta {\bf L}/I$ where 
$I=m_c \sigma_c^2/8$ is the moment of inertia of a disk.
In case of high packing fraction of colloids, it may happen that a single solvent
particle can scatter with several disks in the same timestep $\Delta t_s$. Ignoring
such multiple collisions would cause an attractive
depletion-like force between disks \cite{hech2005}.
This effect can be kept under control allowing a maximum number $N_M$ 
of multiple collisions. It was found empirically that $N_M \simeq 10$ is
the best choice to optimize computational speed and accuracy.
\begin{figure}[t!]
	\centerline{\includegraphics[width=0.98\columnwidth]{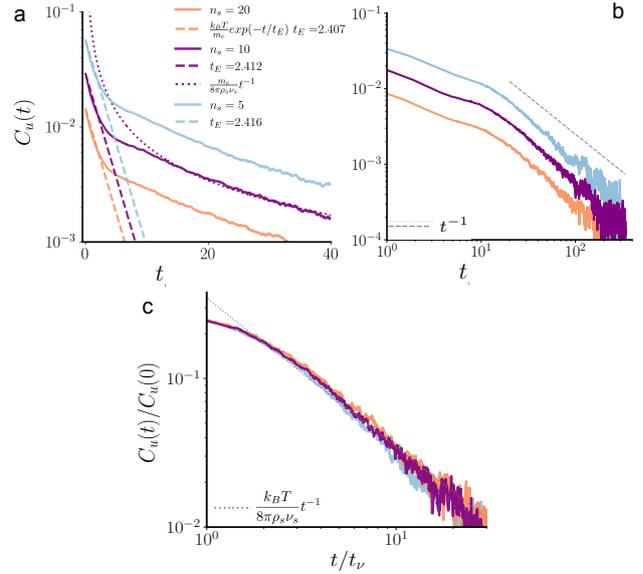}}
	\caption{ \textbf{VACF for different values of the number of solvent particles per cell $n_s$}. Panels (a-c) show the VACF for three different values of $n_s$, namely $n_s=20$ (blue curves) $n_s=10$ (purple curves), and $n_s=5$ (orange curves). For short times (a), the autocorrelation function shows a clear exponential decay, which overlaps well with the theoretical predictions of the the Enskog time,  $t_E$, shown as a dashed line for each case.  At late times (b) simulations show a long time tail $t^{-1}$ (grey dashed lines in panels (b) and (c)\textcolor{black}{, and dotted purple line in panel (a))}. All the data collapse  to the same curve if time is rescaled by $t_{\nu}$ (c). 
	}\label{fig1}
\end{figure}

\begin{figure}[t!]
	\centerline{\includegraphics[width=0.96\columnwidth]{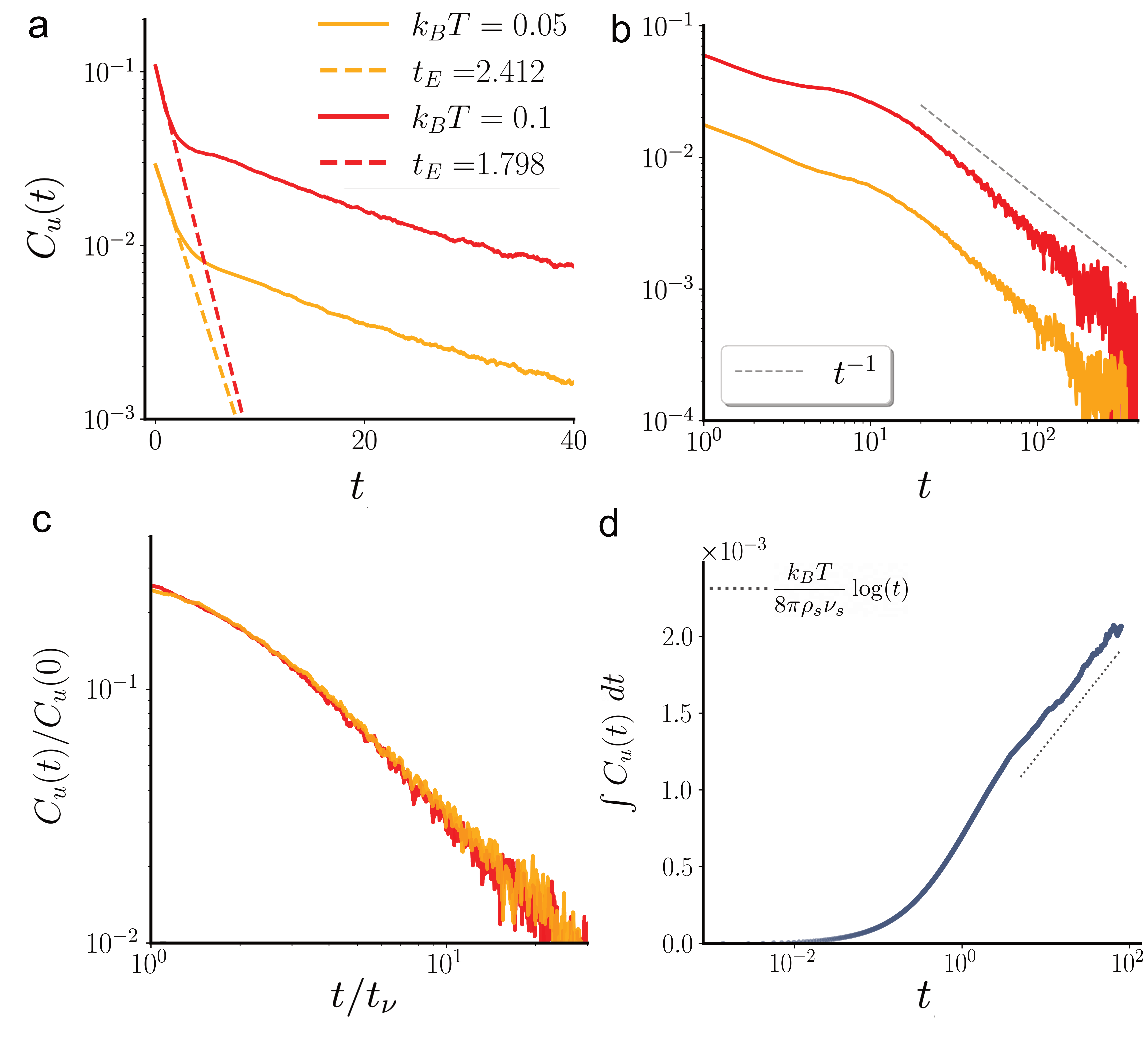}}
	\caption{ \textbf{VACF for different values of the temperature $k_B T$.} 
	Panels (a-c) show the VACF for the values $k_B T=0.05$ (red curves) and 
	$k_B T=0.1$ (yellow curves), for the same number of solvent particles per cell $n_s=10$. For short times (a), the autocorrelation function shows a clear exponential decay, which overlaps well with  the theoretical prediction of the Enskog time $t_E$ shown as dashed line for each case. At late times (b) simulations show a long time tail $t^{-1}$ (dotted line in panel (b)).  All the data collapse  to the same curve if time is rescaled by $t_{\nu}$ (c).  (d) Time evolution, in semi-logarithmic scale, of the diffusion coefficient computed from the integral of the VACF, for the same parameters of the yellow curves of panel (a). The dotted line has the slope $\frac{k_B T}{8\pi\rho_s\nu_s}$.
	}\label{fig2}
\end{figure}
In the case of slip thermal BC, the tangential component of the fluid particle velocity is preserved during the scattering with disks; thus no torque is imparted to colloids. The normal component $v_N$ of the solvent particle new velocity ${\bf v}'$
is sampled from a Gaussian distribution according to the distribution of Eq.~(\ref{pn})
which is centered around the disk velocity ${\bf V}$ (the angular velocity is irrelevant since
collisions are now treated as central) \cite{plipton_colloids_lammps}. 

The choice between no-slip and slip BC is directly connected to the way the axis of colloids ${\bf n}_i$ is evolved. In the first case, the solvent-disk interaction determines directly through torque exchange how colloids diffuse rotationally. In the second case, the rotational diffusion is accounted independently using Eq.(\ref{eq:diffusion}). 
In this paper we choose the slip thermal BC for two reasons. The first one is that in this way we can choose the value of $D_\theta$ independently and match it with the one used in the ABP model. The second reason is that the integration of slip conditions is much faster than no-slip ones, since there is no need of considering the integration of disks angular velocities. %xxxgiusto?
% xxx(the angular velocity is irrelevant since
%collisions are central)xxx.

%, and 
% is particularly suited when the solvent mean free path is  
%much smaller than the disk radius and is able to obtain the correct
%rotational diffusion of a colloid \cite{padd2005,padd2006}.

\textcolor{black}{Since we will mostly deal with non-equilibrium simulations, solvent particles must be coupled to a thermostat to maintain constant temperature. We use the method of locally rescaling
fluid particles velocities $\mathbf{v}_i$ relative to the centre of mass velocity $\bf u$ for each cell by a proper factor that enforces the correct temperature \cite{thermostat}.  
We do not expect that this approach may alter flow profiles since, as later shown, we will adopt a very small cell size $\sigma_s$ compared to variations in flow patterns and a very large
value of $n_s$, the average number of solvent particles per cell.} Note that this implementation ensures only local linear momentum conservation, while angular and total linear momenta are not conserved, as  typically ensured in simulations of swimmers \cite{thee2018}.

%xxxAggiungere parte sul termostato del solvente
%The tstat keyword will thermostat the SRD particles to the specified Tsrd. This is done every N timesteps, during the velocity rotation operation, by rescaling the thermal velocity of particles in each SRD bin to the desired temperature. 
%A random grid shifting is implemented in every collision step to restore the Galilean invariance in a small mean free path limit (λ < σ) [35, 55, 56].

%In the case of no-slip boundary conditions, bounce-back collision rule, which inverts the components of the solvent velocity relative to the surface of collision, could be implemented\cite{lamu2001}. 

\begin{figure*}[t!]
    \centering
    \includegraphics[width=0.96\linewidth]{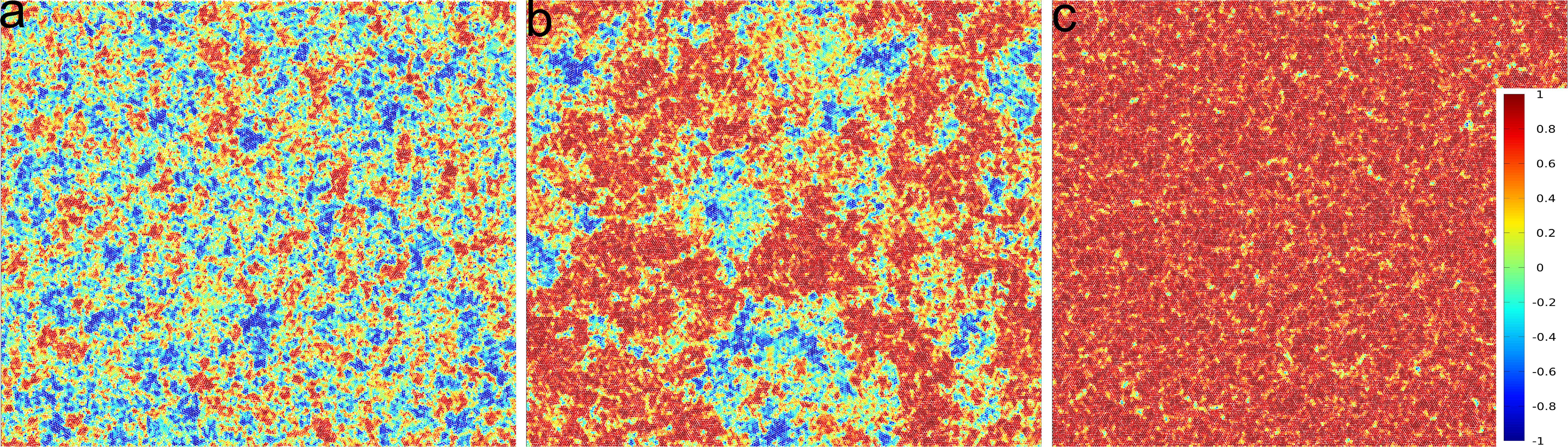}
    \caption{\textbf{Hexatic order parameter color map in the ABP model.} \textbf{(a)-(c)} Color maps of the projection of the local hexatic order parameter of each particle, $\psi_{6,j}$, onto the direction of the system's global average, $\bm{\Psi} = 1/N \sum_j \psi_{6,j}$, at fixed $\textrm{Pe} = 10$ and $m_c = 44$ for $\phi={0.710, 0.730, 0.760}$ respectively, for a system of size $L=256\sigma_c$. 
    %The values of  global hexatic parameter $\psi_6=\frac{1}{N}|\sum_{i}^{N}\psi_{6,i}|$  are $\psi_6=0.015, 0.375, 0.85 $, respectively (see Fig. \ref{fig:hex_phi_time}). }
    }
    \label{fig:langevin_conf}
\end{figure*}

\subsubsection{Parameter choice}
\label{parchoice}

In the case of the MPC fluid, an additional set of simulation parameters has to be set -- $n_s$, $m_s$, $\sigma_s$, $\alpha$, $\Delta t_s$ -- which will be  
 expressed in terms  of the colloids units --  $\tilde{m}$, $\tilde{\sigma}$, $\epsilon$. 
In order to decide the MPC parameter values, a set of criteria, listed below, has to be satisfied.

The first criterion is that the solvent has to behave as a fluid (we remind that MPC particles satisfy an ideal gas equation of state); for such purpose we need to have a Schmidt number $Sc \simeq 10^2-10^3$,  typical of liquids \cite{ripo2004}.
The Schmidt number represents in fact the ratio between 
the rate of momentum diffusion and the rate of mass transfer, and for large values of $Sc$ the dynamics  resembles the one of a liquid  \cite{ripo2005}. 
$Sc$ is defined as $Sc=\nu_s/D_s$.
Values $Sc \sim O(10)$ can be obtained by requiring small values of
$\lambda$  and large
rotation angle \cite{ripo2005}. Note that the choice $\lambda < \sigma_s$ is known to break the Galilean
invariance \cite{ihle2001}, although this problem is  cured by implementing
the random shift procedure \cite{ihle2001} which is here implemented.
By using the expressions of $\nu_s$ and $D_s$ in the limit of $\lambda/\sigma_s \ll 1$, 
we find that the Schmidt number depends only on the mean-free path and takes the simple form\cite{ripo2004}: 
\begin{equation}
Sc \simeq \frac{1}{12 (\lambda/\sigma_s)^2} \ ,
\end{equation}
where the dependence on $n_s$ and $\alpha$ has been omitted since the dominant contribution is with $\lambda$.

The second criterion is that we want to have the same value of the friction $\gamma$ as in ABP simulations, where $\gamma$ has the same role as in the Langevin equation. For the MPC dynamics, this formula is:
\begin{equation}
\gamma= C_{2D}\pi\nu_s\rho_s (\sigma_c/2), 
\end{equation}
where $\rho_s=n_s m_s/\sigma_s^2$ is the solvent density. The coefficient $C_{2D}$ depends on dimensionality \cite{merda2d} and the MPC model considered\cite{theersSRD}.
We performed simulations measuring the velocity of a colloid dragged by a constant force along a direction in 2D and  we fitted a value of $C_{2D}=1.84\pm 0.1$, using six different values of forces and averaging over ten realizations.
It is evident that also the choice of $\gamma$ depends directly only on $\lambda$, when all the other parameters are fixed.

Regarding the active force and the rotational diffusion of the colloids axis, we do not need any change in the parameters chosen for the ABP, as the active force and the rotational diffusion are the same as the ones described in the  equation of motion of the ABP model (equation (\ref{eq:diffusion})). Thus, the Pe number depends only on the colloids parameters, and is already set. 

The last criterion that we need to follow is to have a very low compressibility in  presence of the active force, in order for the fluid to remain homogeneous during the time evolution. This criterion was discussed in \cite{padd2006,theersSRD}. The correct parameters to look at are the Mach number and the Pumping number. The Mach number $Ma$ is given by the ratio between the average fluid velocity $\textrm{v}_s$ due to the external forces (in our case due to activity) and the sound velocity $\textrm{v}_\textrm{sound}=\sqrt{2 k_B T/m_s}$ inside the fluid:
\begin{equation}
Ma=\frac{\mathrm{v_s}}{\mathrm{v}_{sound}}.
\end{equation}
Its value depends directly on flow velocity. In order to reduce compressibility effects of the MPC fluid
it should be $Ma < 0.2$ \cite{lamu2001_1,guyon}.
The Pumping number $Pu$, instead, is the ratio between the active stationary colloid velocity $F_{act}/\gamma$ and the fluid self-diffusion:
\begin{equation}
Pu=\frac{\sigma_c F_{act}}{6 \gamma D_s},
\end{equation}
and should be less than $1$ \cite{theersSRD} in order for the fluid-particle diﬀusion  to be  faster  than activity-induced advection, thus avoiding strong density inhomogeneities in the fluid.

%%%%%%%Vomito i parametri %%%%%%%%%
Following these criteria, we chose  the cell size to be $\sigma_s=0.2\sigma_c$. This guarantees that there is a sufficiently large number of cells covering a colloid  \cite{hech2005}. We fix $\alpha=\pi/2$, $m_s=0.15$ and $n_s=15$ for the fluid. Typically the colloids and solvent mass density should match in order for the colloids to be buoyant, so we set $m_c=44.15$ such that $n_s m_s/\sigma_s^2=4 m_c/(\pi \sigma_c^2)$. This choice provides a 
 good compromise between avoiding compressibility effects\cite{thee2018}, which for example arises if we choose lower $n_s$,  and computational cost, which arises with higher values of $n_s$. We use as $\Delta t_c=10^{-4}$ and  $\Delta t_s=410 \Delta t_c$. The temperature $T$ for the solvent and the other parameters relative to the active force and rotational diffusion remain the same as the one used for ABP. 
These parameters lead to the required values of $\gamma=10.04$ ($\nu_s=0.061$ and $\rho_s=56.24$), $Sc=99.48$, $Ma=0.1$  and $Pu=0.9$ for the highest $Pe=20$ value considered.
We note that the Reynolds number \textcolor{black}{of the fluid} is given by 
\begin{equation}
Re=u_0 \sigma_c / \nu_s=F_{act} \sigma_c / \gamma \nu_s=6 Pu D_s / \nu_s = 6 Pu / Sc ,   
\end{equation}
which is always  much less than one for our choice of the parameters. Thus we are in the low Reynolds number regime.

We start from a close-packed initial configuration of particles positioned in a triangular lattice, forming a slab, and with the orientation of the self-propelled force uniformly distributed. The initial velocities of all particles (fluid particles and colloids) were extracted from a Gaussian distribution with zero mean and variance $k_BT/m_s$ and $k_BT/m_c$ for solvent particles and colloids, respectively. 
Given that  all the MPC and MD parameters are the same we are able  to consider the same exact active colloids system, except for the presence of long range hydrodynamic interactions. 
 We fix the P\'eclet number to $\textit{Pe}=10$ and $20$, and vary the packing fraction $\phi$ between $0.60$ and $0.88$, where the hexatic-liquid transition was found to be critical for $m_c=1$~\cite{nostroDISCHI}. 
For each set of parameters a single realization was considered,  run  between  $10^4$ and $10^5$ simulation time units after steady state was reached, and averaged quantities where performed during this time frame.  To limit the computational cost  for MPC simulations  we always fix the box side to $L=128\sigma_c$, unless otherwise specified.

\subsection{Validation of slip boundary conditions}
%%%%%%%%%%% Slip boundary conditions tests %%%%%
\label{sec:validation}

We focus here on the behaviour of  passive colloids embedded in a solvent to test the accuracy of the previously described slip boundary conditions with respect to known results for 2D hydrodynamics. 
 Following Ref.~\cite{sane2009}, we measure the velocity auto-correlation function (VACF) and the diffusion coefficient $D_c$ of colloids.  The parameters chosen for the simulation are the same as described in the previous section, except that we also varied either  the average number of solvent particles  per cell $n_s$ or the temperature
$k_B T$, always keeping the Schmidt number $Sc \simeq 100$. We considered large systems, with $L=900 \sigma_c$ to  reduce periodic boundary effects. 

\begin{figure}[t]
    \centering
    \includegraphics[width=0.95\columnwidth]{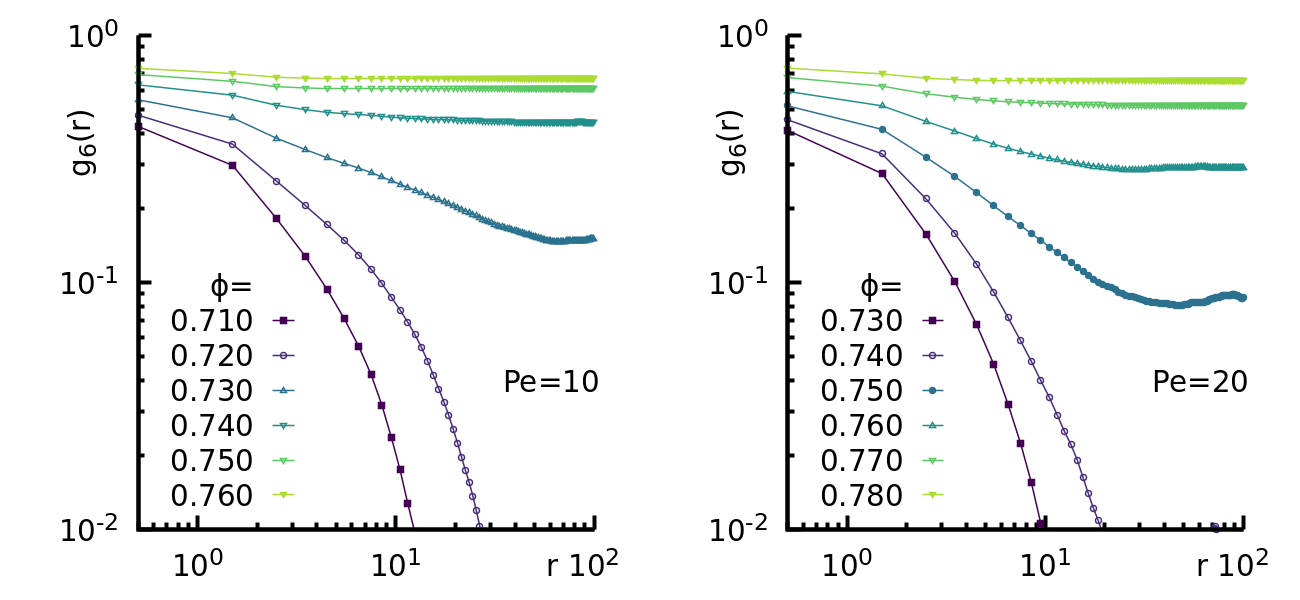}
    \caption{\textbf{Hexatic order correlation functions for the ABP model.}    Hexatic order correlation functions $g_6(r)$ for $m_c=44$ at $\textrm{Pe} = 10$ (left) and $\textrm{Pe} = 20$ (right) for different global packing fractions given in the keys.}
    \label{fig:langevin_corr}
\end{figure}

%\textbf{ATTENZIONE Tutta la derivazione riportata qui e presa da Ref.[12] vale per VACF calcolate lungo una sola direzione. Infatti nelle equazioni 8 e 12 mancherebbe il fattore
%dimensionale 2 al membro di destra. Domanda: le VACF calcolate e plottate nelle nostre figure
%sono calcolate solo lungo una direzione o si e' tenuto conto del fattore 2???}
\begin{figure}[t]
    \includegraphics[width=0.9\columnwidth]{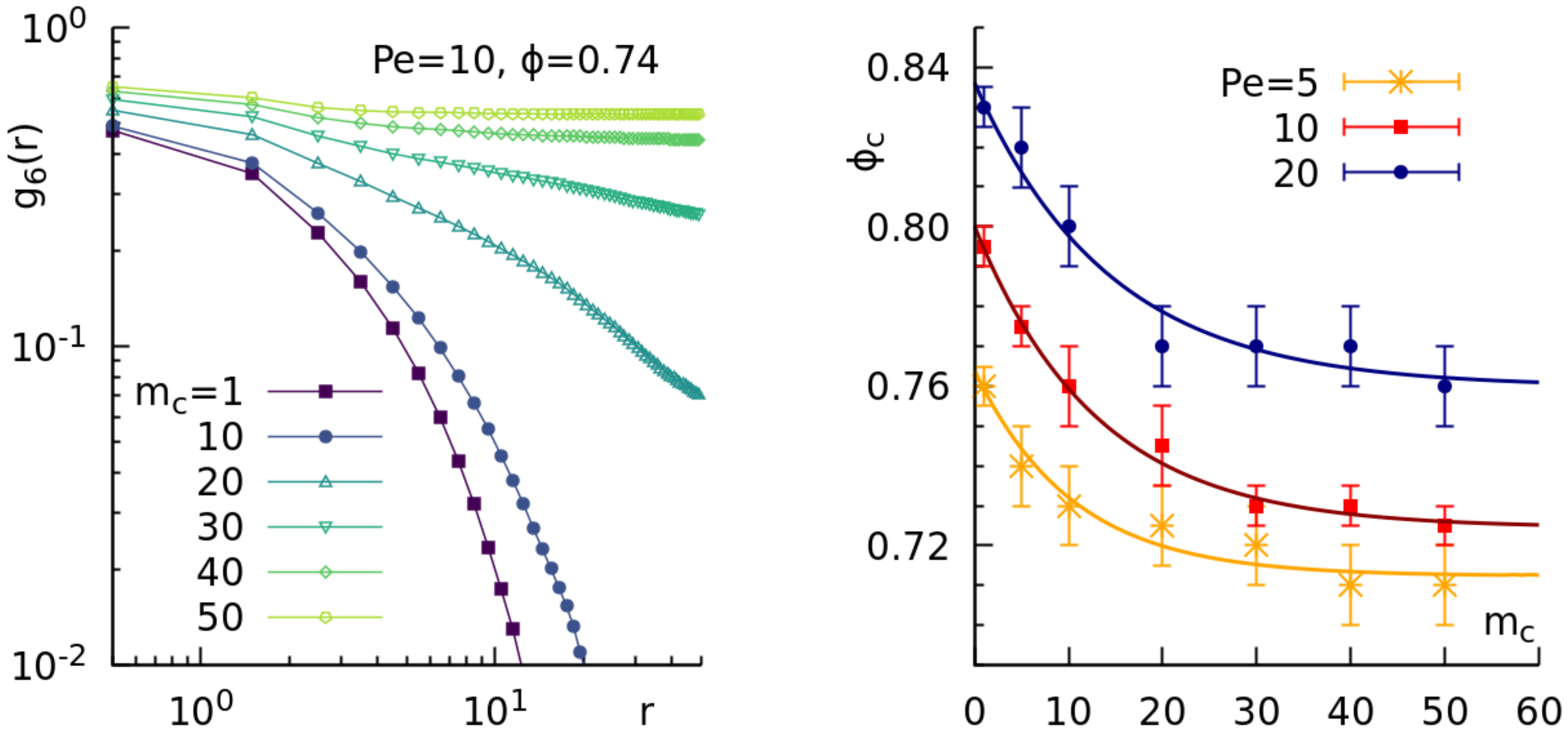}
    \caption{\textbf{ Effects of colloids mass on the Liquid-Hexatic transition}. On the left panel,  orientational correlation functions, $g_6(r)$, at fixed $\rm Pe = 10$ and $\phi = 0.74$ for different values of the mass of the particles given in the keys. On the right panel, the liquid-Hexatic critical density, $\phi_c$, at fixed $\rm Pe = 10$, as a function of the mass of the colloids $m_c$. The solid line is a fit of the data using the function  $\phi_c(m_c)=a+be^{-m_c/c}$, with parameters $a=0.71,\, b=0.05,\, c=10.37$ for $\rm Pe=5$; $a=0.72,\, b=0.08,\, c=12.92$ for $\rm Pe=10$; and $a=0.76,\, b=0.08,\, c=14.30$ for $\rm Pe = 20$.  The error bars correspond to the gap $\Delta\phi$ between the  densities scanned in the simulations for each value of the mass $m_c$.}
    \label{fig:massdep}
\end{figure}
At very short times, when hydrodynamics effects can be neglected, the main contribution to  the overall diffusion  comes from the local random collisions between colloid and solvent particles.  
The VACF is  given by 
\begin{equation}
    C_u(t)=<u(t)u(0)>=\frac{k_B T}{m_c}exp(-t/t_{E}) \ \ ,
\end{equation}
where $u$ is a Cartesian components (either $x$ or $y$) of colloids velocity, $t_{E}=m_c/\xi$ is the Enskog time, that is the  typical velocity decorrelation  time, and $\xi$ the Enskog friction coefficient given in two spatial dimensions by \cite{sane2009} 
\begin{equation} \label{eq:ensk}
    \xi=\frac{3\sqrt{2}}{4}\sigma_{c}n_s\pi^{3/2}\left(k_B T \frac{m_c m_s}{m_c+m_s}\right)^{1/2}\ \ .
\end{equation}
The integral of the VACF is related to the diffusion coefficient $D_c$ through the Green-Kubo relation,
\begin{equation}
D_c=\int_0^{\infty}<u(t)u(0)> dt =\frac{k_B T}{\xi} \ \ .
\end{equation}

However, as well known \cite{alder1970,Ernst1970}  fluid dynamic interactions
have an important effect on the long-time behaviour of the VACF. % as first discovered in MD numerical simulations \cite{alder1970}  and later by analytical calculations \cite{Ernst1970}. 
 Indeed, due to momentum conservation, the asymptotic form of the  VACF shows an algebraic decay of the form 
\begin{equation}\label{eq:long_time_tail}
C_u(t)=\left(\frac{1}{2 \rho_s}\right)\frac{k_B T}{[4\pi(D_c+\nu_s)t]} \ \ ,
\end{equation}
for slip boundary conditions in two dimensions. 
The VACF has a $t^{-1}$ tail, meaning that the diffusion coefficient $D_c$ diverges logarithmically  with time.  The long time tail can be expected to appear on the kinematic time scale $t_{\nu}=\sigma_{c}^2/\nu_s$, that is the time required by the kinematic viscosity $\nu_s$ to diffuse over the colloid radius. 
We validate the slip coupling method introduced in Sec. \ref{sec:solvent_colloids} between solvent and passive colloids by 
testing these predictions.  

Since the kinematic viscosity (\ref{eq:visc})
depends only very weakly on $n_s$, for large values of $n_s$, and given that 
\begin{equation}
C_u(0)=\frac{k_B T}{m_c} \\ ,
\end{equation}
from equipartition, the long-time tails should all scale onto the same curve if time is rescaled by $t_{\nu}$.

Figure \ref{fig1} shows the VACF for three different values of $n_s$ and $k_B T=0.1$. 
As shown in panels (a), for short times,  the autocorrelation function shows clear exponential decay, while at late times 
 (panels (b)-(c)) simulations show a long time tail $t^{-1}$. 
 When plotted as functions of the reduced time $t/t_{\nu}$, all the data collapse onto the same curve (panel (c)). The oscillations visible in panels (b) and (c) for long times originate from sound modes and are a consequence of the finite compressibility of the MPC fluid combined with the periodic boundary conditions \cite{Huang2012}. \textcolor{black}{We checked that this effect decreases increasing the simulation box size.}

 The Enskog friction coefficient (\ref{eq:ensk}) slightly varies with $n_s$;
in order to test the sensibility of the implementation used we fixed $n_s=10$ \begin{figure}[ht!]
    \centering
    \includegraphics[width=0.97\columnwidth]{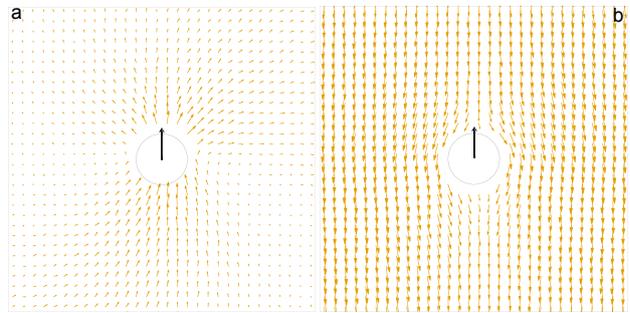}
    \caption{\textbf{Velocity field induced by an active colloid in the AHP model.} \textcolor{black}{Fluid velocity field around an active colloid for $Pe=20$, in the lab frame (a) and in the colloid frame (b). The black arrow indicates the direction of the active force.} 
    }
    \label{fig:neutral_swimmer}
\end{figure}

 and varied the temperature to change the Enskog friction coefficient. 
 Figure \ref{fig2}(a) shows the early time exponential decay of the VACF for the values $k_B T=0.05, 0.1$. The measured values of $t_E$ are in good agreement with the theoretical predictions. Also in this case the long-time tail has the expected $t^{-1}$ slope (panel (b)), and all the curves collapse if time is rescaled by $t/t_{\nu}$ (panel (c)).

\begin{figure*}[htp!]
    \centering
    \includegraphics[width=1.95\columnwidth]{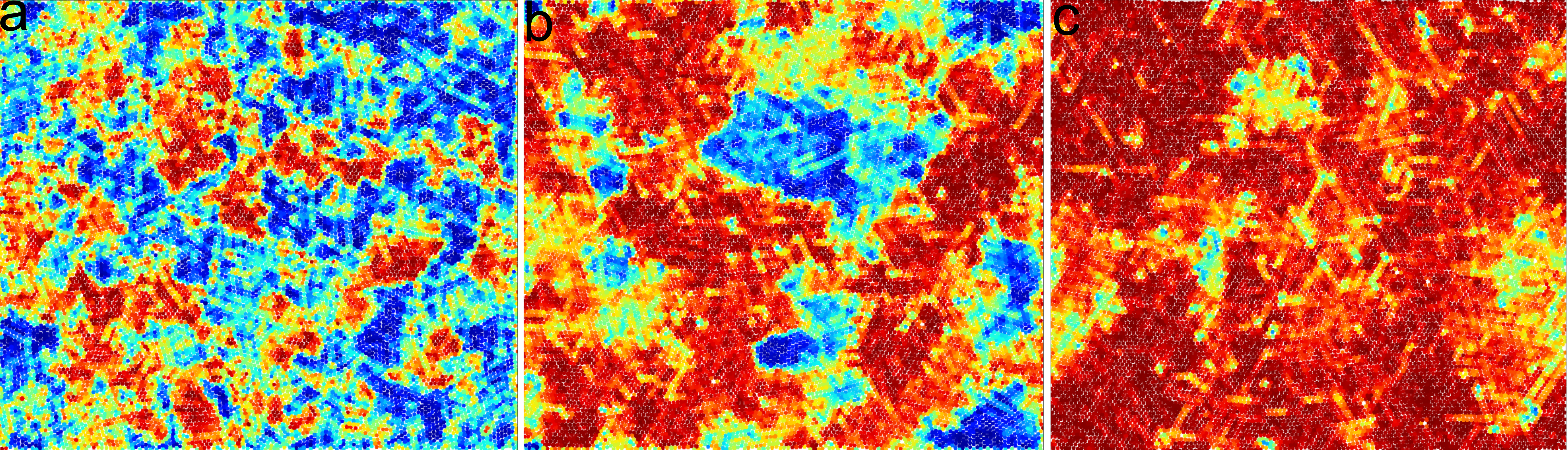}
    \caption{\textbf{Hexatic order parameter color map in the AHP model} \textbf{(a)-(c)} Color maps of the local hexatic order parameter, $\psi_{6,j}$, \textcolor{black}{as reported in Fig.~\ref{fig:langevin_conf}}, for $\textrm{Pe} = 10$ and $m_c=44$, with $\phi={0.78,0.8,0.81}$ from left to right, for a system of size $L=128\sigma_c$. 
    %The values of global hexatic parameter $\psi_6=\frac{1}{N}|\sum_{i}^{N}\psi_{6,i}|$  are $\psi_6=0.07, 0.23, 0.70 $, respectively (see Fig. \ref{fig:hex_phi_time}). }
    }
    \label{fig:hex_conf_srd}
\end{figure*}
Using the Green-Kubo relation, and Eq.~(\ref{eq:long_time_tail}), the diffusion coefficient can be approximated at long times, assuming that $D_c\ll \nu_s$ and that the Enskog and hydrodynamic
contributions to the VACF can be separated, as 
\begin{equation}
\begin{split}
    D_c(t)&=\int_0^{t}<u(t)u(0)>dt \\&\approx\int_{t_{\nu}}^{t}\frac{k_B T}{8\pi\rho_s\nu_s t}\approx \frac{k_B T}{8\pi\rho_s\nu_s}[\ln t]_{t_{\nu}}^{t} \ \ . 
    \end{split}
\end{equation}
  Figure \ref{fig2}(d) shows the temporal evolution of the diffusion coefficient computed from the VACF. On the time scales of the simulation, we observe a behavior consistent with $D_c\simeq ln(t)$, as expected from the $t^{-1}$ tail of the VACF.

\section{Hydrodynamic and variable mass effects on hexatic liquid transition}

In this Section, we discuss the effects of \textcolor{black}{changing the particles mass} for the 2D ABP model  and the role of  hydrodynamics  in the  AHP model, using the numerical framework illustrated in the previous Section.
In particular, we will focus onto characterizing the presence and location of the liquid-hexatic transition, by varying the system density in a region of the phase diagram at intermediate active forces where MIPS does not  occur for over-damped ABP. The latter undergo the transition at $\phi_c=0.795$ for $\textit{Pe}=10$ and at $\phi_c=0.83$ for $\textit{Pe}=20$~\cite{digregorio2017}. 

The transition can be characterized by measuring the hexatic order parameter, $\psi_6(r_i) = \frac{1}{N_i}\sum_{j=1}^{N_i} e^{i6 \theta_{ij}}$, with $N_i$ the number of nearest Voronoi neighbours for particle $i$, and  $\theta_{ij}$ the angle formed between the segment connecting particles $i$ and $j$ and the $x$-axis. 
From $\psi_6(r_i)$ we can compute the hexatic correlation function, defined as:
\begin{equation}
    g_6(r) =\frac{\langle \psi_6(\bm{r}_i)\psi_6(\bm{r}_j) \rangle}{ \langle \psi_6^2(\bm{r}_i) \rangle} \ ,
\end{equation}
where $r = | \bm{r}_i - \bm{r}_j |$. 
The transition between hexatic and liquid phases can be observed by the change in the functional dependence of $g_6(r)$ from exponential decay for short-range order, $g_6(r) \sim e^{-r/l_c}$, where $l_c$ is the correlation length, to algebraic for quasi-long-range order, $g_6(r) \sim r^{-\beta}$.%, with $\beta$ an exponent. % $\beta_c = 1/4$ is the exponent expected from the BKT theory. \textcolor{black}{cosa sono $\alpha$ e $\beta$?} 
%\claudio{non so se conviene specificare il valore di $\beta_c$. A: No, perche poi ce lo chiedono di verificarlo. Servirebbe comunque una ref}.
\textcolor{black}{We use henceforth this criteria to distinguish between the liquid and the hexatic phase in our system.}
In Sec.~\ref{sec:self_sustained} we also discuss from a dynamical perspective how macroscopic flow properties emerge when hydrodynamics is considered.

\subsection{Effects of different colloids mass in the ABP model}
\label{sec:inertia}

%The transition is characterized by a passage from a quasi-long range orientational order, $g_6(r) \propto r^{-\alpha}$ to a short range orientational order, $g_6(r) \propto e^{-r/\xi}$, through the unbinding of topological defects. 

Here we characterize the evolution of ABP following the model  described in Sec~\ref{sec:colloids}  at $\rm Pe = 5, 10, 20$ and compare the results with the ones obtained in Ref.~\cite{digregorio2017}. In particular, while in Ref.~\cite{digregorio2017} only the value $m_c=1$ was considered, here we will study the system with various masses ranging from $m_c=5 \text{ to } 50$. 
Thus, the main difference is that here we are increasing the inertial time $t_I=m_c/\gamma$, ranging from $t_I= 0.5 \text{ to } 5$ while maintaining the persistence time $t_p=1/D_\theta\approx 67$ ~\cite{filyABPs} constant, so that $7\times10^{-3}<pn<7\times 10^{-2}$. 
The use of large masses will allow a direct comparison with the AHP model (where $m_c=44$) that will be used in the following. 
 
We will focus on measuring approximately the value of the critical density $\phi_c$ where the liquid-hexatic transition occurs, computing the hexatic correlation at a fixed Pe within intervals of $\phi$ ranging from 0.05 to 0.1.

%We find that, upon increasing $\phi$, the liquid-hexatic transition occurs both at $\rm Pe = 10, 20$. 
Fig. \ref{fig:langevin_conf}(a)-(c)  show typical configurations at $\rm Pe = 10$,  $m_c=44$ and three different densities. Configurations are colored according to  the local hexatic parameter $\psi_{6,j}$, projected onto its average value. In panel (a) ($\phi=0.71$) we do not observe the appearance of any macroscopic hexatic domain, while in panel c ($\phi=0.76$) we observe a fully hexatically ordered system.
%In panel Fig. \ref{fig:langevin_corr} at fi=xxx si vede una figura dove non si distinguono domini esatici definiti, mentre nel pannello c il sistema e completamente ordinato dal punto di vista orientazionale. 
Panel (b), with $\phi=0.73$, is an intermediate density where  macroscopic and orientationally ordered domains emerge, suggesting that this density is close to  the transition point. 
%sows the mergence fo a coherent hexatic pattern oppure hexatic domanins which suggest that this density is close to  the transition point. 

%The determination of the transition can be done looiking at the behaviour of the correlation funcitons.
In order to locate the liquid-hexatic transition point at a fixed activity, we resort to study the hexatic correlation functions, finding the density at which these functions change from exponential to algebraic decay. Fig.\ref{fig:langevin_corr} shows these functions for $m_c=44$ and $\rm Pe = 10, 20$. At densities below $\phi=0.72$ for $\rm Pe = 10$ and $\phi=0.74$ for $\rm Pe = 20$, we find that the correlations have an exponential decay, while for larger values the behaviours that best fits the decay is that of an algebraic function. Thus, we find that at both activity considered the values where the liquid-hexatic transition occurs are lowered with respect to the ones at $m_c=1$ reported in Ref.~\cite{nostroDIFETTI}, suggesting that the increase in mass enhances the orientational ordering at fixed activity. In particular, we estimate $\phi_c= 0.730\pm0.01$ and $0.760\pm0.01$ for $\rm Pe = 10, 20$, respectively. 

We also checked that this ordering effect occurs while fixing the system density and activity, and increasing the colloids mass. The  correlation functions in Fig. \ref{fig:massdep}, left side, at $\rm Pe = 10$ and $\phi=0.74$, 
 show that by increasing the mass the system  crosses from a liquid state to a hexatic one. We summarize these measurements in the right panel of Fig. \ref{fig:massdep}, where we show the location of the critical density for different Pe and different $m_c$. It is evident that the critical density of the liquid-hexatic transition  continuously decreases increasing the value of the mass for the different Pe considered. Interestingly, the data fit with the function $\phi_c(m_c)=a+be^{-m_c/c}$, with coefficients reported in the caption. 

To summarize, the results showed here point out that an enlarged mass, and therefore an increase in the inertial time $t_I$, has an effect of enhancing the orientational ordering of the system. It is important to note that this is a non-equilibrium effect not present in the passive system. Indeed, we checked (not shown) that in the absence of activity the transition density value is independent of the mass value.%, as there is only one relevant timescale in this case.  
We also observed that the asymptotic values for large $m_c$ (coefficient $a$ in the fitting function) are close to the transition density at $\rm Pe = 0$\cite{digregorio2017}.
When the persistence number is $pn\gtrsim 10^{-2}$ ($m_c\approx10$),   
the system behaves closer to the passive case. On  the other hand, when \textbf{$pn\lesssim 10^{-2}$} the active force has a disordering effect in the hexatic ordering.
\begin{figure}[t!]
    \includegraphics[width=1.0\columnwidth]{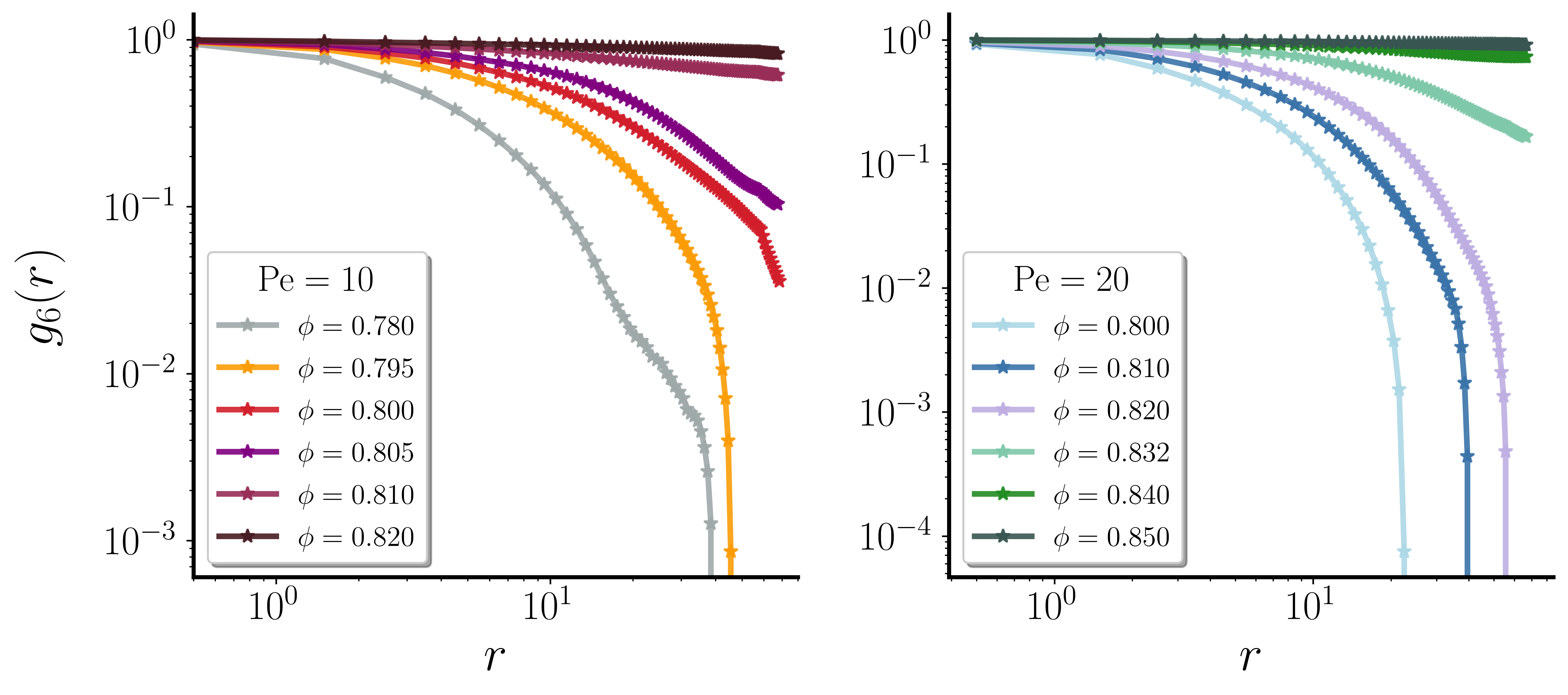}
    \caption{\textbf{Hexatic order correlation function for the AHP model. }  \textbf{(a)-(b)} Hexatic order correlation function $g_6(r)$ at $\textrm{Pe} = 10$ (d) and $\textrm{Pe} = 20$ (e) for different global packing fractions given in the keys  and $m_c=44$.}
    \label{fig:hex_corr_srd}
\end{figure}

\subsection{Hydrodynamics effects}
\label{sec:AHP_results}
We now turn our attention to the role of hydrodynamics by studying the AHP model.  To do so,  we employ the hybrid mesoscopic approach presented and tested in section \ref{sec:model},  where the MPC solvent is coupled with the active colloids to account for hydrodynamic interactions. 
\textcolor{black}{It is important to stress that in our numerical model no tangential flow velocity is imposed to colloids (they are not squirmers), thus the resulting velocity field is the result of collisions between moving colloids and fluid particles. In Fig. \ref{fig:neutral_swimmer} we show the velocity field of our active colloid immersed in a fluid. The flow field strongly resembles that of a neutral swimmer.}

The parameters of AHP, chosen in order to fulfil the constraints discussed in Sec.~\ref{parchoice}, fix the colloid mass to  $m_c=44$ and $\gamma=10$. In this way, the AHP simulation results can be directly compared with the ones of ABP with the same $m_c$. We will scan values of $\phi$ between $0.5$ and $0.85$. 

\subsubsection{Liquid-hexatic transition}

We start by looking at how the ordering properties are affected by hydrodynamics. Fig.\ref{fig:hex_conf_srd}(a)-(c) show, for three different densities at $\textrm{Pe}=10$, the color map of the local hexatic parameter $\psi_{6,j}$ projected onto its average value. In panel (a) ($\phi=0.78$) we do not observe the appearance of macroscopic hexatic domains, but locally we still observe small orientationally ordered regions. These regions appear to become larger upon increasing the density (panel (b), $\phi=0.8$), although global ordering is not observed.  At $\phi=0.81$, panel (c), a single fully hexatically ordered system is observed. Thus, also AHP present a transition between liquid and hexatic phases. 

Fig. \ref{fig:hex_corr_srd} shows the hexatic correlation functions varying the density for $\textrm{Pe}=10, 20$, to be compared with the results presented in Fig. \ref{fig:langevin_corr} for the ABP system. For both values of activity, we find that the hexatic order correlation function shift  from an exponential decay to a power-law         
decay at substantially higher values of packing fraction $\phi$. More precisely the transition is located
at $\phi_c\approx 0.805 \pm 0.01$ for $Pe=10$, and $\phi_c\approx 0.840\pm 0.01$ for $Pe=20$.

\begin{figure}[t!]
    \centering
    \includegraphics[width=0.7\linewidth]{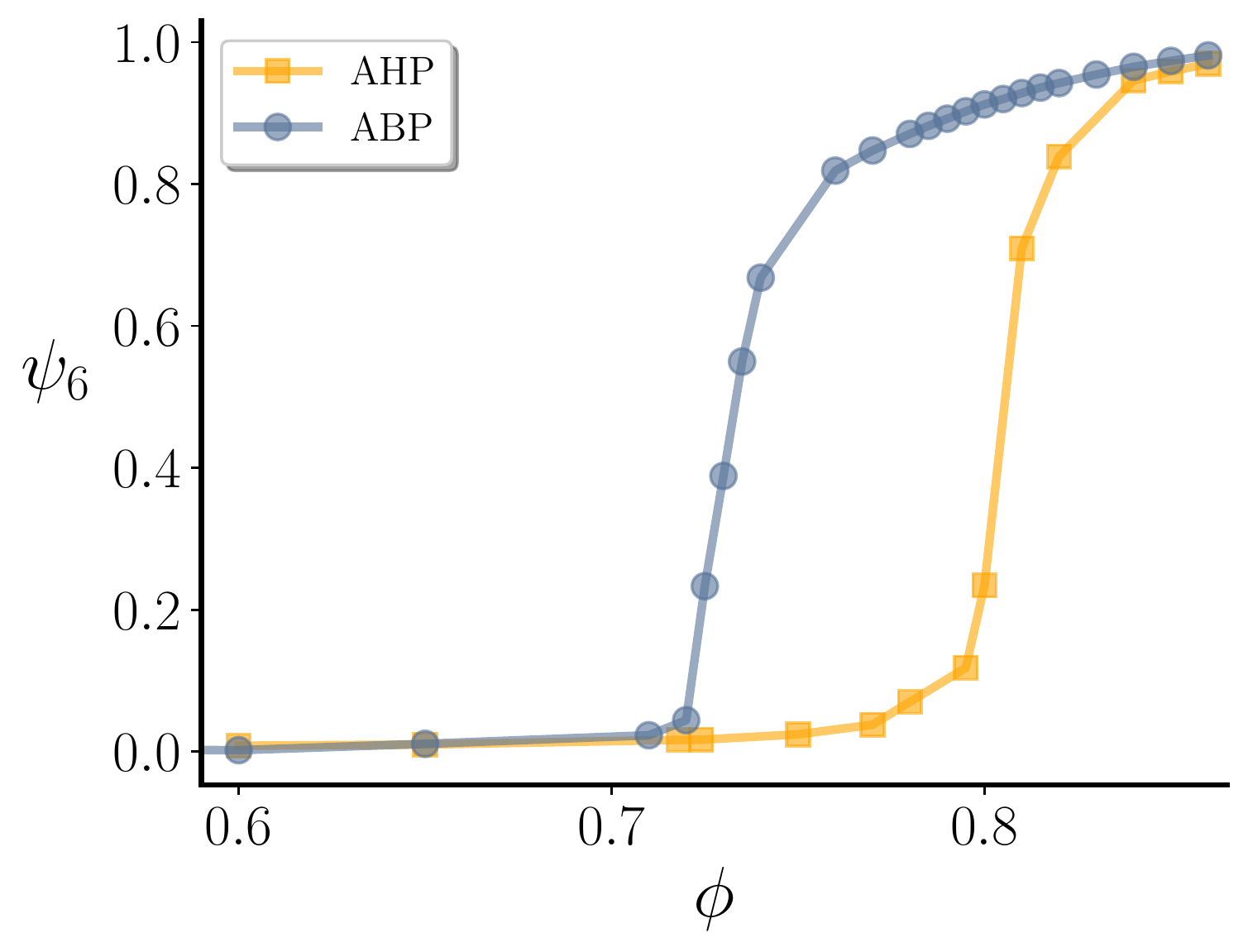}
    \caption{\textbf{Liquid-Hexatic transition  in the ABP and AHP model.}  Global hexatic parameter as a function of the global packing fraction for $\textrm{Pe}=10$ and $m_c=44$,. The orange and blue curves correspond to simulations with and without hydrodynamics, respectively, for active colloids with the same mass. }
    \label{fig:hex_phi_time}
\end{figure}

The increase in value of the transition density $\phi_c$ with respect to the ABP model suggests that the addition of hydrodynamic interactions has a disordering net effect regarding the global orientational order. This is opposite to the effect of \textcolor{black}{increasing the particles mass, which instead promotes hexatic ordering}. Indeed, if we measure the average  global hexatic parameter $\psi_6=\frac{1}{N}|\sum_{i}^{N}\psi_{6,i}|$ as a function of the global packing fraction $\phi$ (Fig.  \ref{fig:hex_phi_time} ), 
%\textcolor{black}{that is another signature of the crossing of the liquid-hexatic transition},
\textcolor{black}{which increases from 0 to 1 as the liquid-hexatic transition is crossed}
we find that the transition is significantly shifted.
Note that both curves converge to almost the same values for very high densities, suggesting that for densely packed systems hydrodynamic does not disrupt the ordering properties of colloids.

%\textcolor{black}{Note that, even if the system sizes for APB and AHP are different, the finite size correction to the value $\phi_c$ is toward lowering this value, thus the difference between APB and AHP is still expected.}
We also checked (not shown) that in the absence of activity, the transition density values that limit the coexistence region of the liquid-hexatic transition of passive colloids~\cite{nostroDISCHI} are not  affected by the presence of hydrodynamic interactions.
However,  hydrodynamics produces other relevant effects which will be now discussed. %play a central role anymore on the ordering properties.  
%\textcolor{black}{It is worth to mention at this point,  that the results so far discussed are in agreement with previous observations regarding hydrodynamic suppression of aggregation phenomena of self propelled particles [Gompper]. GIUSEPPE: not sure this should be here } Non centra quello che stiamo dicendo con la non aggregazione

%Fig. \ref{fig:hex_phi_time} (right panel) shows $|\psi_6|$ over $t$, highlighting a more noisy evolution if HI are taken into account.  

\label{sec:self_sustained}
\begin{figure*}[t!]
 \centering
   \includegraphics[width=1.9\columnwidth]{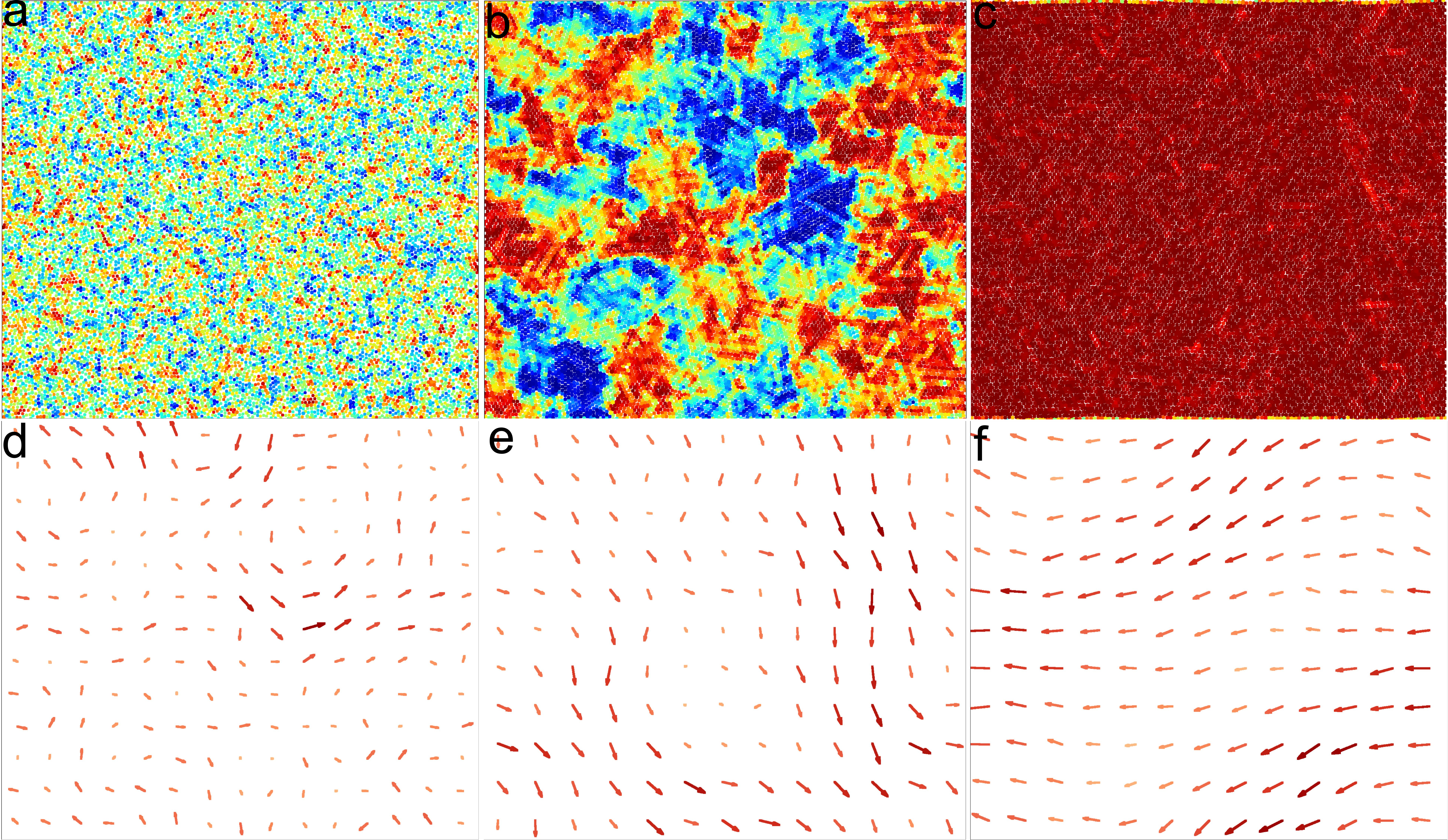}
    \caption{\textbf{Self-sustained active flow.} \textbf{(a)-(c)} Color maps of the local hexatic order parameter for $\textrm{Pe}=10$ and $\phi=0.60, 0.80, 0.86$, respectively. Panels \textbf{(d)-(f)} show the corresponding steady state fluid velocity field. The color code is the same as the one in Fig.~\ref{fig:langevin_conf}.}
\label{fig:vel}
\end{figure*}

\subsubsection{Self sustained motion at high density}

We now want  to better understand the behavior and the role of the fluid velocity field, which for AHP can be locally organized, while the ABP model has no such feature, and rely only on hard-core repulsion.
%need to understand if there are major differences about having or not hydrodynamics explicitly.
%In this regard, the main difference between ABP and AHP is that the latter present a locally organized fluid velocity field, which might play a role in the disordering effect of AHP. ABP model, instead, has no such feature, and rely only on hard-core repulsion, unless additional interactions are introduced, such as the  alignment of active force directions present in the Vicsek-like models\cite{vicsek,levis2018}, which allow particles to move coherently.
Thus, we will have a deeper look into the velocity field of AHP, and if it can trigger a coherent motion of small clusters of particles. 

Fig. \ref{fig:vel} shows the coarse-grained steady state velocity fields of the fluid, $\mathbf{v(r)}$ (panels d-f) along with associated snapshots of the configurations colored according to the hexatic parameter (panels a-c), for AHP with $\textrm{Pe}=10$ and three different values of packing fraction $\phi$. Coarse-grained velocity fields  of the fluid are realized by averaging the velocity of fluid  particles inside blocks of size $4\sigma_c=20\sigma_s$ (such large coarse-graining cells are chosen for the sake of visualization; similar profiles can be obtained with smaller cells). The first density, $\phi=0.60$  (panels (a), (d)), is characterized by the absence of any orientational order. At the same time, however, its corresponding velocity field presents the formation of vortexes along with regions where flow is both not correlated and lower in magnitude. The associated velocity field for the active colloids (not shown) has a matching profile, while the local average direction of the active force is random, and thus not coherent with the velocity field.

Fig.\ref{fig:vel}b,e show instead a larger density  $\phi=0.80$. We observe, here, a case close to the hexatic transition point, with locally formed fluctuating hexatic domains with their typical size remaining stationary over time. Along with these clusters, the flow becomes more coherent than at $\phi=0.60$, with fluid and colloids having again a similar velocity field. Again, we do not observe a local average direction of the active force coherent with the flow field. The same behaviour becomes even more pronounced upon increasing the density ($\phi=0.860$ panels (c) and (f)), where the system is fully orientationally ordered. In this case, the associated flow field becomes an unidirected self-sustained flow, with particles moving typically on the same direction, and with the global direction of the flow slowly changing over time. Interestingly, this behaviour is similar to  travelling bands occurring in Vicsek-like models \cite{vicsek,levis2018}, where an additional alignment interaction  of active force directions is introduced, which allows particles to move coherently. Velocity correlations between particles have also been found in systems of ABP with different persistence times~\cite{capriniPRL,caprini2020,Bettolo_Marconi_2021,capriniINERTIASM},  flowing crystals made of spontaneously aligning self-propelled hard disks \cite{douchot} and self-sustained spontaneous flows in active gels \cite{review2019,softmatter,giordano2021,negro2020,NEGRO_2019}. 

We do not have at the moment a full theoretical understanding of the emergence of the coherent motion, which occurs even when there is no orientational ordering. We can only try to interpret the phenomenology in the following way: the self-propulsion force of colloids continuously injects energy into the fluid, setting it into motion. Fluid particles can later self-organize their motion in a coherent form, and drag colloids along their direction of motion, which is not necessarily the same direction of the active force of each particle.%, as in the Vicsek model, but in the direction of the locally organized fluid field itself. 

\begin{figure}[t!]
    \centering
    \includegraphics[width=0.95\columnwidth]{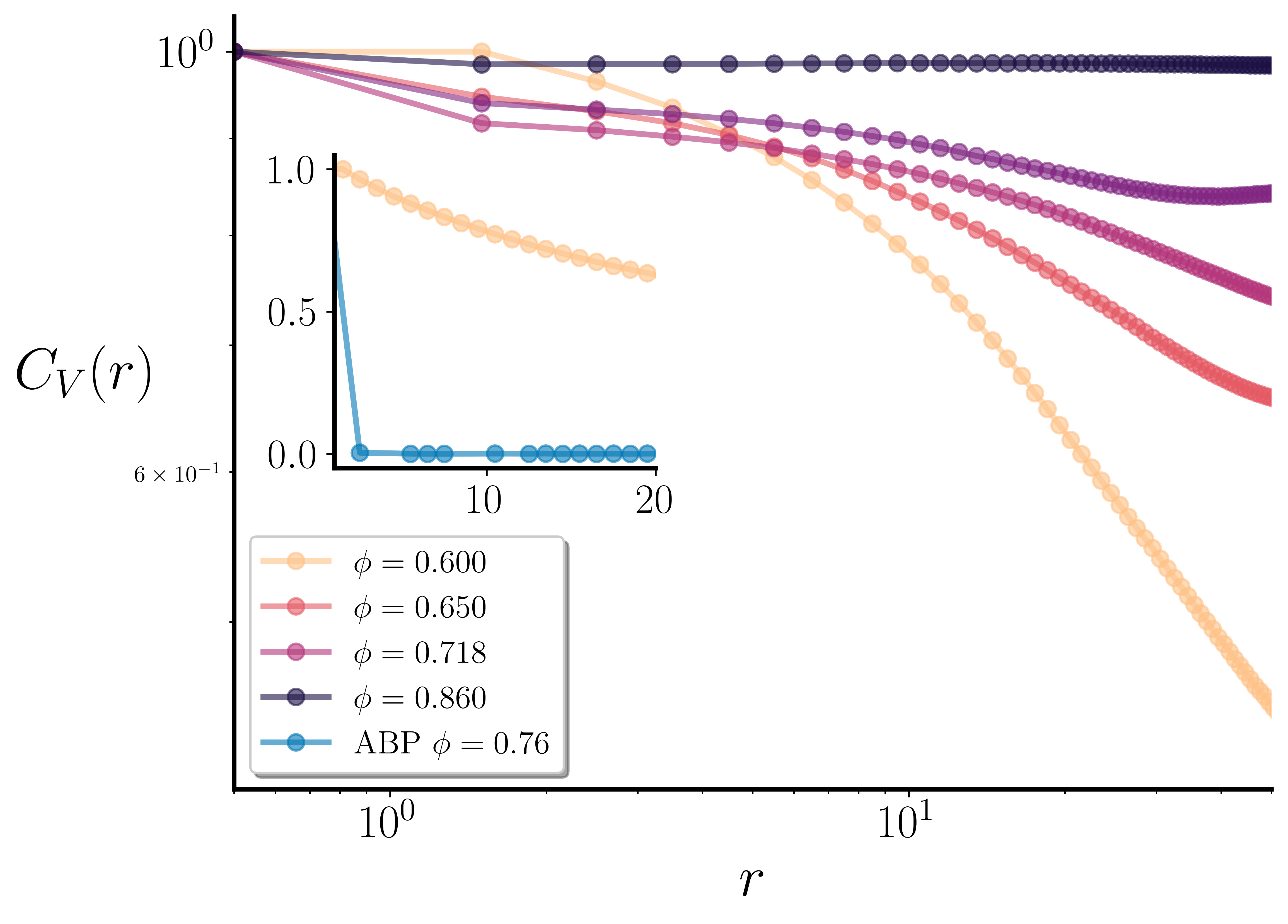}
    \caption{\textbf{Spatial velocity correlations.} Spatial velocity correlation functions $C_{\textrm{v}}(r)$, for different values of $\phi$, for $\textrm{Pe}=10$. In the inset the velocity correlation function of ABP at $Pe=10$ and $\phi=0.760$ is compared with $\phi=0.60$ for AHP at the same $Pe$. }
    \label{fig:spatial_vel_corr}
\end{figure}

\begin{figure}[t!]
    \centering
    \includegraphics[width=0.95\columnwidth]{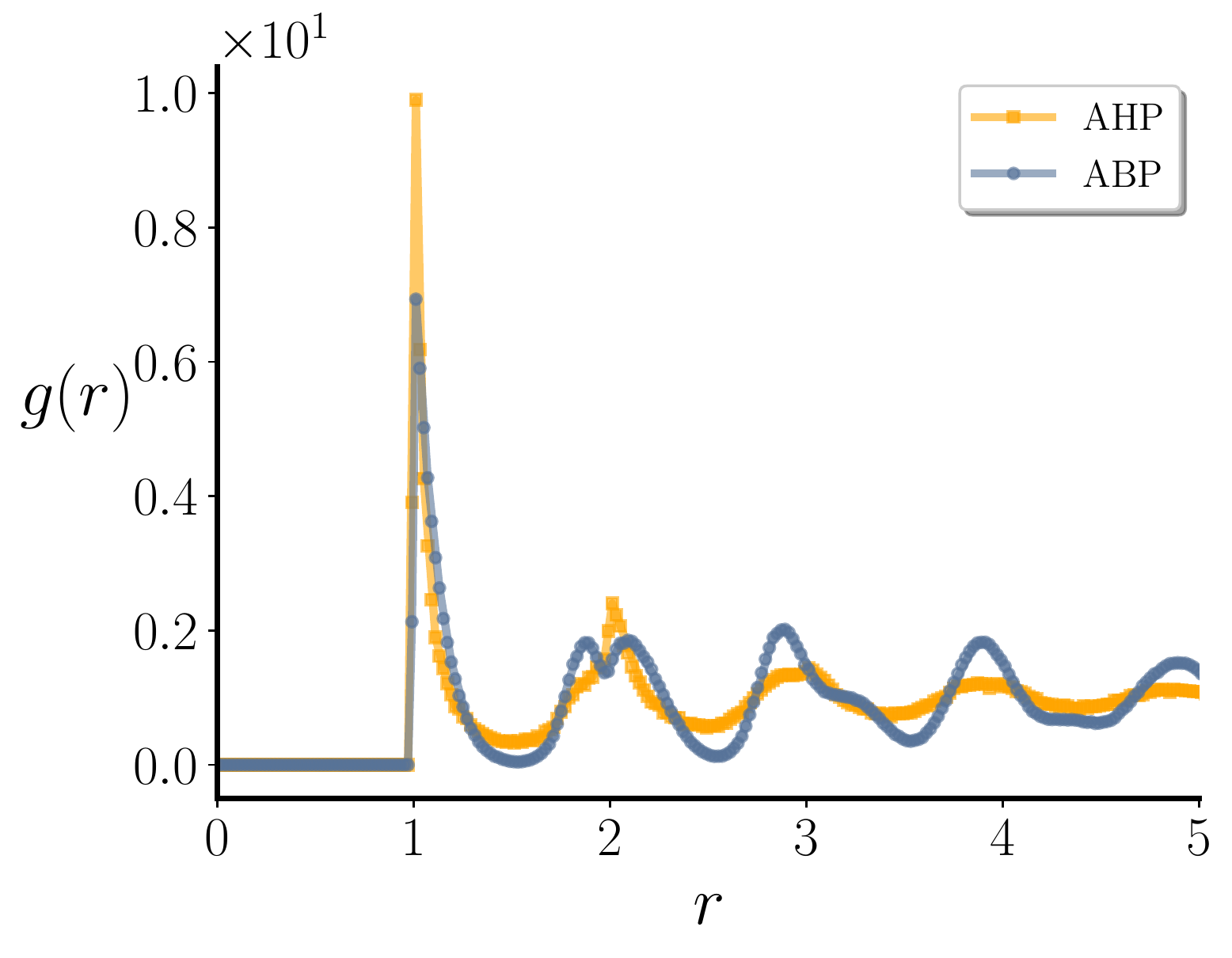}
    \caption{\textbf{Radial distribution function.} \textcolor{black}{Radial distribution function for ABP and AHP models with $\phi=0.750$ at $Pe=10$.}}
    \label{fig:radial_dist}
\end{figure}

A quantitative measure of this transition to  unidirected self-sustained flow, as a synergetic effect of self-propulsion and hydrodynamic interactions,   can be obtained by measuring the spatial velocity correlation function for the fluid velocity: 

\begin{equation}
    C_{\textrm{v}}(\mathbf{r})=\frac{\langle\mathbf{v(r)v(0)}\rangle}{\langle\mathbf{v(0)}^2\rangle} \ \ . 
\end{equation}
 Fig. \ref{fig:spatial_vel_corr} shows $C_{\textrm{v}}(\mathbf{r})$ for different values of $\phi$, for $Pe=10$. For low values of $\phi$ (see e.g. $\phi=0.600$) the curve shows an exponential decay. This corresponds to the case shown in Fig. \ref{fig:vel}(d), characterized by the presence of isolated vortexes. When we increase the density, we observe that the velocity correlation has a slower decay, or a longer correlation length. %After a certain
 Above density $\phi\simeq 0.730$, the correlation becomes  almost constant.
We note that the transition in the velocity correlations between exponential and algebraic decay does not manifest itself at the liquid-hexatic transition, since the latter appears  at higher values of $\phi$. In the inset of Fig. \ref{fig:spatial_vel_corr} the velocity correlation function for ABP in the hexatic phase ($Pe=10$ and $\phi=0.760$) is also shown for comparison. It shortly decays to zero, while for AHP, even in the liquid case (yellow curve), the decay is much slower.
\textcolor{black}{To gain more insights on the effects of hydrodynamic interactions we report the radial distribution function $g(r)$ in Fig. \ref{fig:radial_dist}, for $\phi=0.750$ both for ABP and AHP at $Pe=10$.  We observe that the presence of fluid in AHP does not considerably change the position of the peaks in the radial distribution function. However, we notice that the intensity of the peaks is  enhanced in the ABP case, meaning that the fluid interferes with ordering, thus shifting the hexatic transition to higher densities. }

\begin{figure}[htp!]
    \includegraphics[width=0.94\columnwidth]{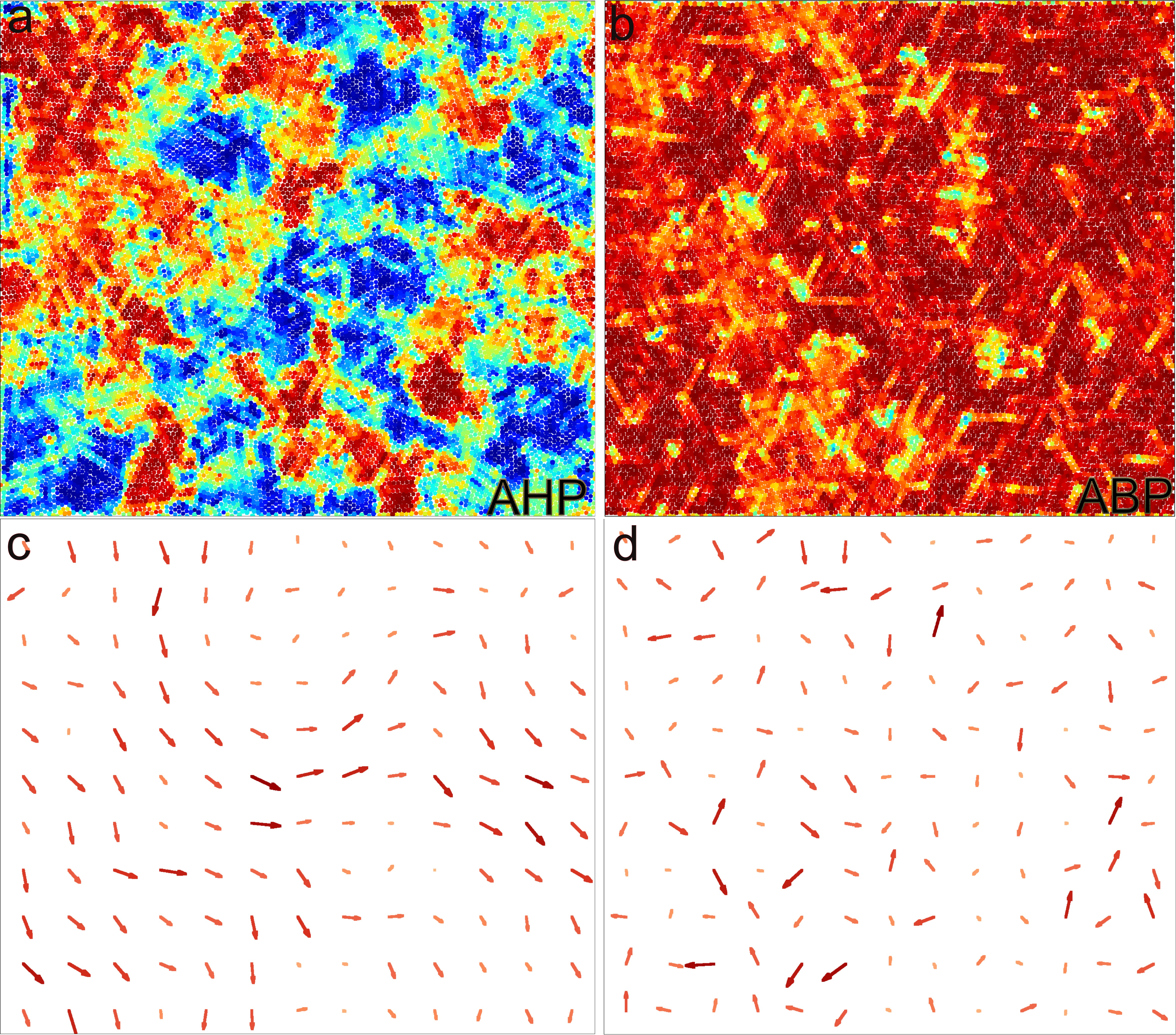}
    \caption{\textbf{Switch between ABP and AHP model.} (a)-(b) Snapshots of the system at different times $t= 10^4, 10^5$  of the local hexatic order parameter, at $\textrm{Pe}=10$ and $\phi=0.795$, before (panel (a)) and after (panel (b)) switching off hydrodynamics. The corresponding colloids velocity fields are shown in panels (c)-(d). }
    \label{fig:time_snaps}
\end{figure}

As a last check, we switched off/on  hydrodynamics by just removing/adding the solvent particles and adding/removing the Langevin friction and noise terms in the colloids equation of motion (\ref{eq:diffusion}). This enables us to check if a stationary AHP configuration is naturally able to relax to a stationary conformation of the ABP when hydrodynamics is switched off. 
We choose $Pe=10$ and $\phi=0.795$, a density 
%which
where the system 
is hexatically disordered/ordered with/without  hydrodynamics. The results are shown in Fig. \ref{fig:time_snaps}.
We start with AHP in a fully ordered configuration; after an equilibration time of $10^4$ simulation time units, the system forms fluctuating ordered domains which change over time but do not grow in size (panel (a)). We then turn off hydrodynamics, and the system gradually sets after $t=10^5$ simulation time units to a almost fully hexatically ordered conformation (panel (b)).  The corresponding colloids velocity field is shown in panels (c)-(d).  Note that the configuration is still not fully ordered  only due to the large time required to relax to the fully ordered state; however we observe that the global hexatic parameter is steadily growing over time. Switching on hydrodynamics again the system returns to the configuration shown in Fig. \ref{fig:time_snaps}(a).

\section{Conclusions}
We have studied with extensive simulations the role of \textcolor{black}{particles mass}  and hydrodynamics in active colloids, and showed how they affect the liquid-hexatic transition  in an intermediate activity regime in which MIPS does not occur yet ( $\textit{Pe}=10, 20$). 

We have first characterized the ABP by changing their mass, while maintaining the same Pe and $D_{\theta}$, so that we have a non-trivial interplay between the inertial time and the persistence time $t_p=1/D_{\theta}$.  We showed that the critical density of the transition  is shifted to a lower density upon increasing the colloid mass. This critical density is close to the one found at $\textit{Pe}=0$, suggesting that inertia has an orientational ordering effect on the system bringing the system closer to equilibrium behaviour and counteracting the disordering role of self-propulsion. 

When hydrodynamic interactions are taken into account, we found instead that the liquid-hexatic transition moves towards higher values of packing fraction $\phi$, thus suggesting that hydrodynamics has a net effect of orientationally disordering the system. %This is compatible with the results presented in \cite{theersSRD}, where it is argued that MIPS  is in fact inhibited by hydrodynamics.
We also analyzed the fluid velocity field of AHP, and found at $Pe=10$ two results: i) the formation below $\phi\approx 0.72$ of small correlated velocity field regions, characterized by the presence of vortices, that are not associated to any local orientational ordering; ii) the arisal above $\phi\approx 0.720$ of a self sustained motion, with the fluid particles  moving in one direction.
 This change in behavior has been characterized by measuring the spatial velocity correlation which was found to change from an exponential to an algebraic decay.

Regarding the role of inertia, it will be interesting in the future to reconstruct a phase diagram similar to the one of Ref.\cite{nostroDISCHI}, by characterizing in more detail the hexatic phase and the location of the solid phase.
Regarding AHP, instead, it will be necessary to better describe  the physical mechanisms producing the vortices at smaller densities and the transition to a self-sustained motion at larger densities.
It remains an open question whether such a scenario is still encountered in quasi 2D and 3D geometries as well as in experiments with wet active colloids.
It would also be of interest to investigate the effect of no-slip boundary conditions, which would completely determine the colloid angular diffusion and could induce additional cooperative effects,  the effects of changing colloidal mass, and investigating in more details particle-particle flow interactions and local velocity field effects.%On this regard,  recent  experiments on self-propelled polar disks \cite{douchot} have shown that polar and ordered structures organizes into a coherent sheared flow, which is made possible by the localization of shear along intermittent stacking faults. These observations are also well reproduced by a SPP model where velocity is coupled to the disks polarity....
 We hope that our results can boost further research in this direction. 

\section*{Acknowledgements}

This work was possible thank to the access to Bari ReCaS e-Infrastructure funded by MIUR through PON Research and Competitiveness 2007-2013 Call 254 Action I and MARCONI at CINECA (Project INF22-fieldturb) under CINECA-INFN agreement.  We acknowledge funding from MIUR Project No. PRIN 2020/PFCXPE. 

\section*{Author contribution statement}

All authors contributed equally to conceptualizing the research, analysing the results and writing the paper. GN and CBC carried out the simulations. 

\section*{Data Availability Statement}

The datasets generated during and/or analysed during the current study are available from the corresponding author on reasonable request.

% \textcolor{blue}{We don't find any sign of coexistence between the two phases (as in the case of  as in the case of passive discs at Pe=0 ). Measuring the pressure would also distinguish between first order and critical behaviour, but this would require a much larger set of simulations especially in the case of AHP. xxxcheck con Claudio, opppure conclusioni }

\bibliographystyle{unsrt}
\bibliography{biblio}
\end{document}